\documentclass[10pt]{iopart}

\usepackage{tabls}
\usepackage{url}
\usepackage{amsbsy}
\usepackage{graphicx}
\usepackage{color}
\usepackage{hyperref}

\newcommand{\eqref}[1]{{(\ref{#1})}}

\begin{document}

\title[Non-sky-averaged sensitivity curves for space-based GW observatories]{Non-sky-averaged sensitivity curves for space-based gravitational-wave observatories}

\author{Michele Vallisneri and Chad R.~Galley }
\address{Jet Propulsion Laboratory, California Institute of Technology, Pasadena, CA 91109}


\begin{abstract}
The signal-to-noise ratio (SNR) is used in gravitational-wave observations as the basic figure of merit for detection confidence and, together with the Fisher matrix, for the amount of physical information that can be extracted from a detected signal. SNRs are usually computed from a sensitivity curve, which describes the gravitational-wave amplitude needed by a monochromatic source of given frequency to achieve a threshold SNR.
Although the term ``sensitivity'' is used loosely to refer to the detector's noise spectral density, the two quantities are not the same: the sensitivity includes also the frequency- and orientation-dependent response of the detector to gravitational waves, and takes into account the duration of observation.
For interferometric space-based detectors similar to LISA, which are sensitive to long-lived signals and have constantly changing position and orientation,
exact SNRs need to be computed on a source-by-source basis.
For convenience, most authors prefer to work with sky-averaged sensitivities, accepting inaccurate SNRs for individual sources and giving up control over the statistical distribution of SNRs for source populations.
In this paper, we describe a straightforward end-to-end recipe to compute the \emph{non}-sky-averaged sensitivity of interferometric space-based detectors of any geometry. This recipe includes the effects of spacecraft motion and of seasonal variations in the partially subtracted confusion foreground from Galactic binaries, and it can be used to generate a sampling distribution of sensitivities for a given source population. In effect, we derive \emph{error bars} for the sky-averaged sensitivity curve, which provide a stringent statistical interpretation for previously unqualified statements about sky-averaged SNRs.
As a worked-out example, we consider isotropic and Galactic-disk populations of monochromatic sources, as observed with the ``classic LISA'' configuration. We confirm that the (standard) inverse-rms average sensitivity for the isotropic population remains the same whether or not the LISA orbits are included in the computation. However, detector motion tightens the distribution of sensitivities, so for 50\% of sources the sensitivity is within 30\% of its average. For the Galactic-disk population, the average and the distribution of the sensitivity for a moving detector turn out to be similar to the isotropic case. 
\end{abstract}


\maketitle

\section{Introduction}

The new window on the Universe promised by gravitational-wave (GW) astronomy is opening, but we cannot as yet see much outside. The eagerly awaited direct detection of GWs is likely to be achieved in the next few years with second-generation ground-based interferometric detectors such as Advanced LIGO and Virgo \cite{2010CQGra..27h4006H,avirgo}, but at the moment we still operate in a predetection era when the planning, design, and configuration of future GW observatories is a major research topic. For instance, third-generation ground-based instruments such as the Einstein Telescope \cite{grg43} promise much greater reach and yield, but their detailed design is still unsettled. In the space-based detection effort, after the ESA--NASA collaboration on LISA \cite{lisasciencecase,Jennrich:2009p1398} was halted in 2011, ESA has been studying a slightly descoped European-led mission (known as NGO, or eLISA \cite{esastudy,amaldiproc}) for possible launch at the end of the decade; in late 2011 the NASA Physics of the Cosmos program issued a Request For Information (RFI, \cite{nasarfi}) on possible low-cost designs of space-based GW observatories in case NGO is not selected, but with a farther implementation horizon.

The GW \emph{sensitivity curve} of proposed detector designs describes the GW amplitude needed by monochromatic sources to be detectable with a prescribed statistical confidence within a given observation time; it is arguably the primary metric of the detectors' scientific value, since it figures in the computation of signal-to-noise ratios (SNRs, see Sec.\ \ref{sec:definitions}), detection rates, and parameter-estimation accuracies, even for chirping sources. However, because space-based detectors will move and rotate with respect to the long-lived sources that inhabit the low-frequency GW spectrum, there is \emph{no such thing as a single sensitivity curve}---the sensitivity is different for sources with different geometries, and even different timing. Due to the motion of the Earth, this is true also for ground-based detectors and continuous sources such as deformed pulsars. For the sake of convenience, GW scientists have typically resorted to using sky-, \mbox{polarization-,} and inclination-averaged sensitivity curves, with the understanding that the SNRs quoted for any individual source would be inaccurate, and would usually be provided without an error estimate. (While in the GW literature the sensitivity is sometimes conflated with the noise spectral density of the detector, these quantities have different units and represent different concepts. However, we shall see below that the sensitivity can be used to define the \emph{effective strain noise} of the detector, which incorporates the detector's GW response.)

In this paper we advocate the general approach of computing the geometry-dependent sensitivity over a large, physically motivated sample of geometries, and then using the resulting distribution of sensitivities to characterize the error, as it were, of sky-averaged sensitivity curves. As an example, we work out the ``classic LISA'' sensitivity to monochromatic binary sources that are distributed isotropically over the sky or across a realistic Galactic disk. This analysis shows, for instance, that for less than 5\% of source geometries the sensitivity is worse than twice\footnote{Since the sensitivity is defined as the source strength needed to achieve a given SNR, here larger is worse, smaller is better.} (or better than 2/3) the average sensitivity, if we use the ``inverse rms'' definition of average discussed in Sec.\ \ref{sec:definitions}; and that for 50\% of source geometries the sensitivity is within $\pm 30\%$ of the average: see Figs.\ \ref{fig:sensitivities1} and \ref{fig:sensitivities2}. We also confirm that for an isotropic source population the average sensitivity computed, as is standard practice, by neglecting the orbital motion of the detector coincides closely with the average sensitivity computed by including the motion; for a Galactic-disk population the stationary-detector average sensitivity is different, but the moving-detector sensitivity follows closely the isotropic result. 

To enable this analysis, we provide a straightforward, end-to-end recipe to compute the exact source-geometry--dependent sensitivity of space-based interferometric detectors similar to LISA, including NGO and most of the mission concepts submitted to the NASA RFI \cite{nasarfi}. Indeed, this paper can be considered as a useful companion to those studies. The recipe is by no means original: it finds its origins in the work of Armstrong, Estabrook, and Tinto \cite{2000PhRvD..62d2002E,whitepaper}, it is analogous to the work of other authors \cite{2000PhRvD..62f2001L,2003PhRvD..67b2001C}, and it expands on the formalism introduced in \cite{2008CQGra..25f5005V}. However, the algorithm described in this paper is self confined and ready to use, and it is presented in the context of a careful formal definition of sensitivity.

This paper is organized as follows. Section \ref{sec:definitions} is an in-depth discussion of the formal and practical definition of sensitivity and of its possible averages; Sections \ref{sec:noise} and \ref{sec:signal} introduce the necessary ingredients for our sensitivity recipe (namely, the noise and GW response of LISA-like detectors, including heterodox examples such as ``bow-tie'' and ``syzygy'' configurations); Section \ref{sec:sensitivity} puts these ingredients together to compute the distribution and average of the classic-LISA sensitivity to monochromatic sources (distributed isotropically and across a Galactic disk), and presents sensitivity plots, tables, and fits. Last, \ref{app:motion} describes the simple model of the LISA orbits used in Sec.\ \ref{sec:sensitivity}, while \ref{app:snr2fit} provides an accurate fit to the average classic-LISA sensitivity.

\section{On a general definition of sensitivity}
\label{sec:definitions}

In signal-processing applications, the detection signal-to-noise ratio (SNR) is defined loosely as the strength of a signal divided by the root-mean-square detector noise in the same set of data. Intuitively, the SNR is a metric of the confidence, quality, and information content of a detection. In the context of a detection scheme, we can make precise statements to back up this intuition.

\subsection{Matched filtering}
For instance, in the matched-filtering detection of a signal $h$ immersed in Gaussian additive noise $n$, the SNR of the data $s = h + n$ \emph{after filtering} by a \emph{normalized} signal template $\hat{t}$ is given by
\begin{equation}
\rho(s;\hat{t}) =
4 \, \mathrm{Re} \int_0^{\infty} \frac{s^*(f) \hat{t}(f)}{S_n(f)} \,\mathrm{d}f
\label{eq:snr}
\end{equation}
(with the conventions of \cite{1994PhRvD..49.2658C}, but our notation), where $s(f)$ and $\hat{t}(f)$ are the Fourier transforms of the detector data and of the template $\hat{t}$, respectively; ``${}^*$'' denotes the complex conjugate; and $S_n(f)$ is the one-sided power spectral density of the noise, defined by the ensemble average $\langle n^*(f) n(f') \rangle = S_n(f) \delta(f-f') / 2$ for $f>0$. A template $t$ is normalized when $\rho(t;t) = 1$.

With this definition, the \emph{false-alarm probability} of $\rho(s;\hat{t})$ exceeding a chosen detection threshold $\rho_\mathrm{thr}$ \emph{for noise alone} scales as $\exp(-\rho_\mathrm{thr}^2/2)$, so higher $\rho(s;\hat{t})$ translates into greater detection confidence of a signal $h \propto \hat{t}$. 
Indeed, the \textit{a priori} detectability of a signal $h$ is characterized by its \emph{optimal SNR} $\sqrt{\rho(h;h)} \equiv \mathrm{SNR}_\mathrm{opt}(h)$, which is the SNR that would be obtained, on average over noise realizations, using a perfectly matching template.

\subsection{Sensitivity and horizon distance}
These considerations lead to a natural definition of detector \emph{sensitivity} to a given class of signals as the \emph{source strength} that yields optimal SNR equal to a fiducial detection threshold $\mathrm{SNR}_\mathrm{thr}$. The definition of source strength is somewhat conventional: 
in matched-filtering searches for GW signals (the concern of this paper), we can write $h(t) = A \, h_0(t)$, isolating an amplitude parameter, and then define
\begin{equation}
\mathrm{sensitivity}(h_0) =
\left\{\begin{array}{c}
\mbox{$A$ needed for $\mathrm{SNR}_\mathrm{opt}$} \\
\mbox{to equal $\mathrm{SNR}_\mathrm{thr}$}
\end{array}\right\}
= \frac{\mathrm{SNR}_\mathrm{thr}}{\sqrt{\rho(h_0;h_0)}}.
\label{eq:h0sens}
\end{equation}
Because $S_n$ appears at the denominator in Eq.\ \eqref{eq:snr}, this definition of sensitivity is in effect rms noise divided by rms signal ($h_0$) of fiducial strength.
Thus, stronger noise makes for worse (numerically higher) sensitivity, which requires stronger signals to achieve the same detection quality.
Sensitivity can be interpreted readily in terms of the \emph{horizon distance} out to which detection is possible: we write $h = (d_0/d) A_0 h_0$ (with $d$ the luminosity distance and $A_0$ the amplitude at distance $d_0$), and define
\begin{equation}
\mathrm{horizon}(h_0) = \frac{d_0 A_0 \sqrt{\rho(h_0;h_0)}}{\mathrm{SNR}_\mathrm{thr}}.
\end{equation}

For GW applications, a useful quantity to compute is the \emph{sensitivity to monochromatic sources} [with signals $h(t) = A \, \cos(2 \pi f_0 t + \phi_0)$], which is related to $S_n(f_0)$ through Eqs.\ \eqref{eq:h0sens} and \eqref{eq:snr} by 
\begin{equation}
\mathrm{sensitivity}(f_0) = \mathrm{SNR}_\mathrm{thr} \frac{\sqrt{S_n(f_0)}}{\sqrt{T}},
\label{eq:fsens}
\end{equation}
with $T$ the time of observation. (The equation is dimensionally correct since $[S_n] = \mathrm{Hz}^{-1}$.) This relation highlights two important (if well known) facts: first, the SNR for continuous, non-chirping sources grows linearly with the square root of $T$; second, the sensitivity to monochromatic sources contains the same information as $S_n(f)$, and therefore can be used to reconstruct the SNR of generic sources through Eq.\ \eqref{eq:snr}.

\subsection{Detector response}
This straightforward idealized picture still requires some elaboration to reflect actual GW detectors. This is because the signals recorded by GW detectors are not simply proportional to the GW strains at the detector locations; rather, they are \emph{functionals} of the strains that reflect a variety of physical effects:
\begin{enumerate}
\item Detector geometry yields the characteristic \emph{antenna patterns} that describe detector response to GWs of given polarization and direction of propagation; 
\item Detectors may need \emph{calibration}---the empirical measurement of the proportionality and lag of detector output vs.\ GW strain, as a function of frequency;
\item Detectors may move (with the Earth, or in space) significantly over the timescale of GW signals;
\item For wavelengths comparable to (or smaller than) the extent of the detector, characteristic time signatures appear in the response.
\end{enumerate}
This list is not all inclusive. With sufficient generality for our purposes, we are going to assume that the detector output $s_\mathrm{det}$ is described (in Fourier space) by
\begin{equation}
s_\mathrm{det}(f) = \mathsf{R}(f) : \mathsf{h}(f) + n_\mathrm{det}(f),
\end{equation}
where the sans-serif quantities denote symmetric trace-free tensors and ``:'' their inner products, $\mathsf{R}(f)$ is the detector's \emph{response function}, and $n_\mathrm{det}$ is the instrument noise. In the case of ground-based detectors, $\mathsf{R}(f)$ includes the detector's calibration function \cite{Abadie2010223}, while the product $\mathsf{R}(f) : \mathsf{h}(f)$ produces the well-known detector antenna patterns $F_+$ and $F_\times$ \cite[Sec.\ 7.2]{Maggiore:1900zz}. In the case of space-based interferometric detectors, $\mathsf{R}(f)$ is a complex function that encodes the time-delay structure of the so-called {\it Time-Delay Interferometry} (TDI) observables (see Sec.\ \ref{sec:noise} below), and which is also a function of the position and orientation of the spacecraft at the time of observation.

\subsection{Averaged sensitivity}
It is useful to distinguish between the dependence of SNR (and therefore sensitivity) on the \emph{intrinsic} source parameters $\theta^\mu$ (such as binary masses and spins) that determine the time evolution and structure of the GW signal, and its dependence on the \emph{extrinsic} parameters $\Omega^\mu$ (such as the source's sky position and luminosity distance) that describe, as it were, the contingent presentation of the signal to the detector. Now, the dependence of sensitivity on the extrinsic parameters is arguably less interesting;\footnote{Once we detect a source, we will certainly wish to learn about its sky position (e.g., to look for electromagnetic counterparts); but we would not normally decide \emph{a priori} to look preferentially for sources from a given sky region.} furthermore, we wish to define sensitivity as a function of as few parameters as possible, so that we may plot it or fit it easily. Thus, it is often desirable to average the sensitivity over the extrinsic parameters. We can, for instance, rms-average the optimal SNR over $\Omega$,  
\begin{equation}
\bigl\langle \mathrm{SNR}_\mathrm{opt}(\mathsf{h}_0) \bigr\rangle_{\Omega,\mathrm{rms}} = \sqrt{ \left\langle
4 \int_0^\mathrm{\infty} \frac{\bigl|\mathsf{R}(f) : \mathsf{h}_0(f)\bigr|^2}{S_\mathrm{det}(f)} \, \mathrm{d}f \right\rangle_\Omega 
} \; ,
\end{equation}
and then define the averaged sensitivity as $\mathrm{SNR}_\mathrm{thr} / \langle \mathrm{SNR}_\mathrm{opt}(\mathsf{h}_0) \rangle_{\Omega,\mathrm{rms}}$.
Indeed, if the GWs can be written as $\mathsf{h}(f;\Omega,\theta) = \mathsf{e}(\Omega) \, h(f;\theta)$, with $\mathsf{e}(\Omega)$ a constant polarization tensor (possibly complex), and $h(f;\theta)$ a function, we can exchange the $\Omega$-averaging with the integration over frequency,
\begin{eqnarray} \fl
\bigl\langle \mathrm{SNR}_\mathrm{opt} (\mathsf{h}_0) \bigr\rangle_{\Omega,\mathrm{rms}}
&=& \sqrt{
4 \int_0^\mathrm{\infty} \! \left\langle\frac{
\bigl|\mathsf{R}(f) : \mathsf{e}\bigr|^2 \times \bigr|h_0(f)\bigr|^2}{S_\mathrm{det}(f)}\right\rangle_{\!\Omega} \! \mathrm{d}f
} \\ &=&
\sqrt{
4 \int_0^\mathrm{\infty} \! \left\langle \frac{\bigr|h_0(f)\bigr|^2}{S_\mathrm{eff}(f,\Omega)}
\right\rangle_{\!\Omega} \!\mathrm{d}f
} = \sqrt{4 \int_0^\mathrm{\infty} \frac{
|h_0(f)|^2}{S^\mathrm{rms}_\mathrm{eff}(f)} \, \mathrm{d}f}, \nonumber
\end{eqnarray}
where we have folded the detector's response and $\Omega$ dependence into the noise,
yielding the $\Omega$-dependent \emph{effective strain noise} $S_\mathrm{eff}(f,\Omega) = S_\mathrm{det}(f) / |\mathsf{R}(f) : \mathsf{e}|^2$, and the (inverse-rms) averaged effective strain noise $S_\mathrm{eff}^\mathrm{rms}(f)$, which can be used to define a sensitivity to monochromatic sources according to Eq.\ \eqref{eq:fsens}.

An important example where the factorization $\mathsf{h}(f;\Omega,\theta) = \mathsf{e}(\Omega) \, h(f;\theta)$ takes place is the dominant quadrupolar emission from inspiraling binaries. For monochromatic binaries, for instance, we have
$\mathsf{h} = \mathsf{e}_+(\Omega) h_+(t) + \mathsf{e}_\times(\Omega) h_\times(t)$ with
\begin{equation} \fl
\qquad h_+      = A (1 + \cos^2 \iota) \cos(2 \pi f_0 t + \phi_0), \quad
h_\times = - 2 A \cos \iota     \sin(2 \pi f_0 t + \phi_0),
\label{eq:mono1}
\end{equation}
and therefore (for $f > 0$)
\begin{equation}
\mathsf{h}(f) = \bigl[(1 + \cos^2 \iota) \mathsf{e}_+ + \mathrm{i} (-2 \cos \iota) \mathsf{e}_\times\bigr] \times \frac{A}{2} \mathrm{e}^{i \phi_0} \delta(f - f_0).
\label{eq:mono2}
\end{equation}
Here $f_0$ is the GW frequency and $A = \frac{2}{d_L} \bigl(\frac{G \mathcal{M}_c}{c^2}\bigr)^{5/3} \bigl(\frac{\pi f_0}{c}\bigr)^{2/3}$, with $d_L$ the luminosity distance, and $\mathcal{M}_c = \mu^{3/5} M^{2/5} = (m_1 m_2)^{3/5}/(m_1 + m_2)^{1/5}$ the \emph{chirp mass} (defined in turn by the binary component masses $m_1$ and $m_2$, or by the total mass $M$ and reduced mass $\mu$). We should emphasize that this definition of $A$ is conventional; another possibility is to take the signal strength as the inclination- (and possibly time-) averaged $h^2 = h_+^2 + h_\times^2$ at the Solar-system barycenter (SSB). This is useful because $h^2$ follows from the binary's energy loss.

Possible alternative definitions of averaged sensitivity, which do not allow the definition of the equivalent of $S^\mathrm{rms}_\mathrm{eff}(f)$, include the inverse of mean SNR [i.e., $\mathrm{SNR}_\mathrm{thr} / \langle \mathrm{SNR}_\mathrm{opt} \rangle_{\Omega}$] and
the mean of inverse SNR [i.e., $\mathrm{SNR}_\mathrm{thr} \times \langle \mathrm{SNR}^{-1}_\mathrm{opt} \rangle_{\Omega}$]. If we are using the horizon distance to compute an expected detection rate that is proportional to horizon \emph{volume}, then the relevant average is
$\langle \mathrm{SNR}^3_\mathrm{opt} \rangle_\Omega^{1/3}$ (which Thorne \cite{thorne.k:1987} suggests can be approximated as $(3/2) \langle \mathrm{SNR}^2_\mathrm{opt} \rangle_\Omega^{1/2}$ for interferometric detectors).


\section{Measurement noise of space-based GW interferometers}
\label{sec:noise}
\begin{figure}
\includegraphics[width=\textwidth]{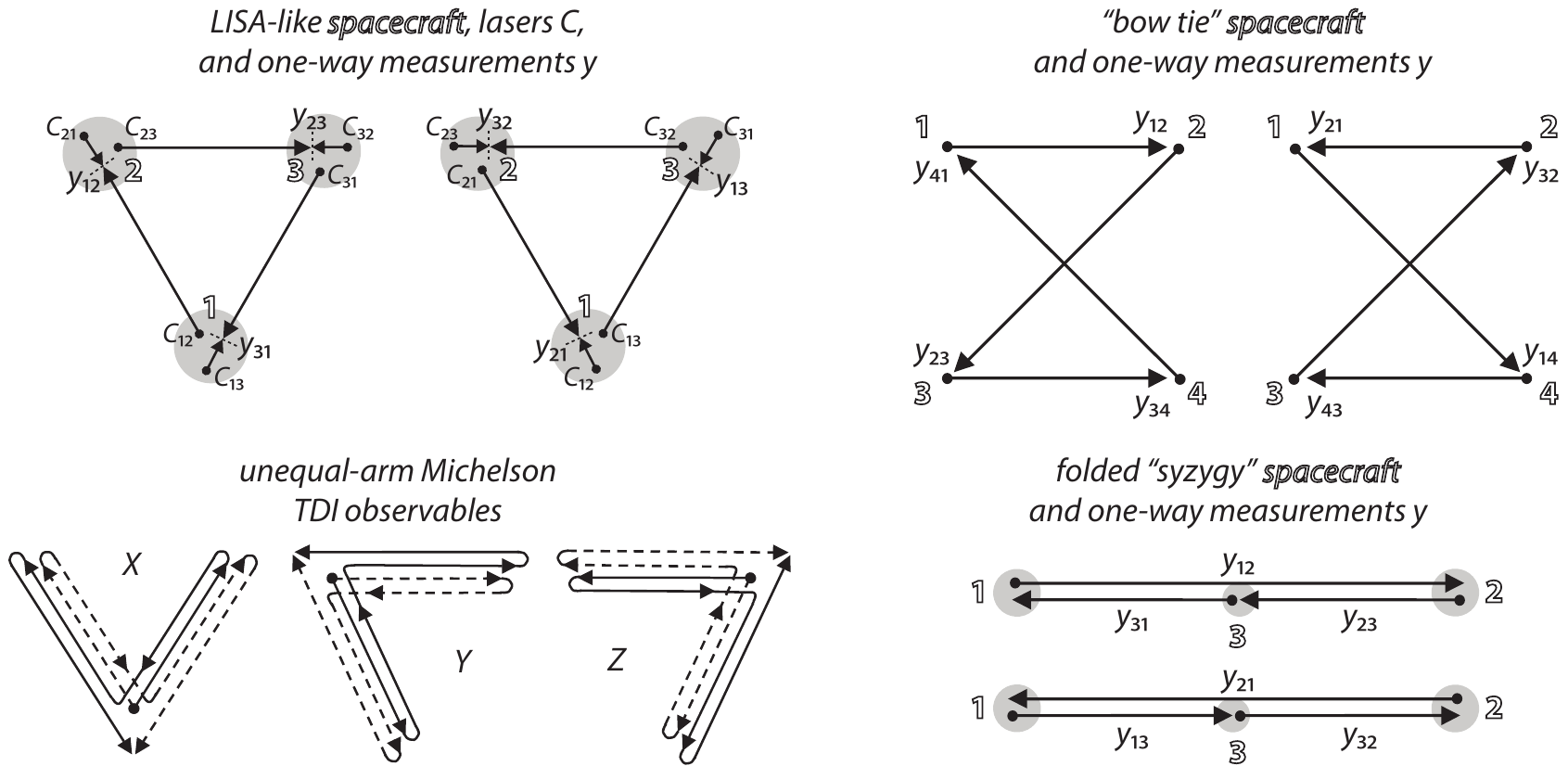}
\caption{\textbf{Left, top}: inter-spacecraft measurements $y_{ij}$ in a LISA-like three-spacecraft
configuration. Each arrow represents a laser $C_{ij}$ transmitted between spacecraft; laser frequencies are compared in the $y_{ij}$ measurements, taken at the dotted lines. (We do not show the referencing of the measurement to the freely falling test masses.)
\textbf{Left, bottom}: the unequal-arm Michelson observables of time-delay interferometry, centered around spacecraft 1 (observable $X$), 2 ($Y$), and 3 ($Z$). TDI observables can be seen as \emph{synthesized interferometers}, because the sequence of
time-delayed measurements reproduces the path of light in a table-top interferometer, with consecutive measurements (e.g., $y_{21}(t) + y_{12}(t - L_{21})$) playing the role of mirrors, and arithmetic subtraction substituting for beam splitters and interferometric recombination at photodetectors \cite{2005PhRvD..72d2003V}.
\textbf{Right, top}: inter-spacecraft measurements $y_{ij}$ in a four-spacecraft, eight-link ``bow tie'' configuration. (For clarity, we do not show the local comparison lasers for each measurement.)
\textbf{Right, bottom}: inter-spacecraft measurements $y_{ij}$ in a folded ``syzygy'' configuration \cite{2003PhRvD..68f2001E}, considered under the name \emph{SGO lowest} in the 2011 NASA concept study \cite{nasarfi} (note however that the $y_{23}$ and $y_{32}$ links would not be implemented in that mission concept).
\label{fig:tdi}}
\end{figure}

Space-based GW observatories similar to LISA \cite{lisasciencecase} comprise at least three spacecraft orbiting the Sun in formation, with $\sim 10^6$ km inter-spacecraft distances (for LISA, the formation is a quasi-equilateral triangle, with sides $\sim 5 \times 10^6$ km).
GW strain is measured by monitoring the oscillating distance between freely-falling test masses contained by the spacecraft, using heterodyne laser interferometry. In what is known as ``drag-free'' operation, the spacecraft hover around the test masses to protect them from external perturbations, performing slight orbital corrections with very precise micro-Newton thrusters. The lasers exchanged by the spacecraft are neither split nor reflected, as they would be in a laboratory interferometer. Rather, dual \emph{one-way} laser links are established between pairs of spacecraft $(i,j)$, and each link provides a test-mass--referenced interferometric measurement $y_{ij}$ of the phase (or equivalently, the frequency) of the laser incoming from spacecraft $i$, compared to spacecraft $j$'s local, outgoing laser. Thus, the notations $y_{ij}$ and $y_{ji}$ refer to independent measurements that compare the same two lasers on different spacecraft (see top left panel of Fig.\ \ref{fig:tdi}).

While the GW response of these measurements is of course central to the computation of sensitivity, we shall first discuss their noise, which dictates the structure of the TDI observables of LISA-like observatories (\cite{1999PhRvD..59j2003T}; see \cite{2005LRR.....8....4D,2005PhRvD..72d2003V} for extensive reviews).
For simplicity, in what follows we gloss over several resolvable complications that are discussed at length in the literature,\footnote{Examples of complications are the cancellation of ultrastable-oscillator (USO) noise, the motions of optical benches, the offsets and drifts of laser frequencies, and the Doppler beat notes due to relative spacecraft motion.} we refer our formalism to the ``classic'' non-\textit{strapdown} LISA architecture,\footnote{In the classic architecture, the incoming laser is reflected off the test mass and then compared to the local laser on the optical bench; in a strapdown configuration, the incoming laser is compared directly to local laser, and a separate interferometric measurement is made of the position of the test mass with respect to the optical bench.} and we use the simplest possible notation (cf.\ Table I of \cite{synthlisa}).

\subsection{Noise in one-way and corner-spacecraft measurements}
The total noise in the individual $y_{ij}$ measurements, written in terms of fractional frequency fluctuations, is then given by
\begin{equation}
y^\mathrm{noise}_{ij}(t) = 
C_{ij}\bigr(t - L_{ij}(t)\bigl) - C_{ji}(t) + y^\mathrm{op}_{ij}(t) - 2 y^\mathrm{pm}_{ij}(t):
\label{eq:ynoisesimple}
\end{equation}
for this measurement, which takes place at time $t$ on spacecraft $j$:
\begin{itemize}
\item $C_{ij}$ represents the frequency noise for the laser sent from spacecraft $i$ to $j$, which is also compared locally with the laser $C_{ji}$ sent from spacecraft $j$ to $i$;
\item $L_{ij}$ is the one-way transit time of light along the link, which delays the instantaneous noise of the incoming laser, as measured in $y_{ij}$;
\item $y^\mathrm{pm}_{ij}$ is the line-of-sight velocity fluctuation of the test mass on spacecraft $j$;
\item $y^\mathrm{op}_{ij}$ is a sum of optical-path and photon-shot noise for the heterodyne measurement.
\end{itemize}
``Corner'' spacecraft, which participate in laser links with two other spacecraft, include two measurement systems, each consisting of a laser, test mass, and optical bench. On such a spacecraft (say, $j$), two additional measurements $z_{ij}$ and $z_{kj}$ are made to compare the frequency of the two lasers and to remove the noise of the optical fiber that connects the two optical benches. These measurements have total noise given by
\begin{eqnarray}
z^\mathrm{noise}_{ij}(t) &=& C_{kj}(t) - C_{ij}(t) + 2 y^\mathrm{pm}_{kj}(t), \label{eq:znoisesimple} \\
z^\mathrm{noise}_{kj}(t) &=& C_{ij}(t) - C_{kj}(t) + 2 y^\mathrm{pm}_{ij}(t). \nonumber
\end{eqnarray}

\subsection{Time-delay interferometry}
The $y_{ij}$ measurements cannot be used as such to search for GWs, because the laser frequency noise of space-worthy lasers would greatly overwhelm all other noise sources, as well as the expected response to GWs. Instead, the science data will consist of \emph{TDI observables}: linear combinations of several $y$ and $z$, appropriately time delayed so that all instances of laser frequency noise $C$ cancel out, or rather are suppressed by several orders of magnitude. For instance, the \emph{unequal-arm Michelson} TDI observable $X$ (which, as the name suggests, exactly cancels laser noise in configurations with constant but unequal arms) is given by
\newcommand{\del}[1]{\mathcal{D}_{{#1}}}
\begin{eqnarray}
\label{eq:unequalmich}
\fl \quad X(t) = {} & \bigl[ y_{31}(t) + \del{31} y_{13}(t) + \del{13} \del{31} y_{21}(t) + \del{21} \del{13} \del{31} y_{12}(t) \bigr] \\
\fl & - \bigl[ y_{21}(t) + \del{21} y_{12}(t) + \del{12} \del{21} y_{31}(t) + \del{31} \del{12} \del{21} y_{13}(t) \bigr] \nonumber \\
\fl & + \bigl[-\del{13} \del{31} \del{12} \del{21} z_{31}(t) + \del{12} \del{21} z_{31}(t) + \del{13} \del{31} z_{31}(t) - z_{31}(t) \bigr]/2 \nonumber \\
\fl & + \bigl[\del{12} \del{21} \del{13} \del{31} z_{21}(t) - \del{12} \del{21} z_{21}(t) - \del{13} \del{31} z_{21}(t) + z_{21}(t) ) \bigr]/2, \nonumber
\end{eqnarray}
where we have introduced time-delay operators $\mathcal{D}_{ij}$ such that for any measurement $x(t)$,
\begin{equation}
\fl \quad \mathcal{D}_{ij} x(t) \equiv x\big(t - L_{ij}(t)\big), \;\;
\mathcal{D}_{kl} \mathcal{D}_{ij} x(t) \equiv x\Big(t - L_{ij}(t) - L_{kl}\big(t - L_{ij}(t)\big)\Big), \;\; \ldots
\label{eq:delays}
\end{equation}
Note that in general the delays $L_{ij}(t)$ and $L_{ji}(t)$ in the two directions along the same arm are not equal, because of relative spacecraft motion.

TDI observables can be seen as \emph{synthesized} interferometers, because their $y_{ij}$ components trace the paths and reproduce the phase delays of light across recognizable interferometric configurations, such as a Michelson interferometer for $X$---see the bottom left panel of Fig.\ \ref{fig:tdi}.
Now, the time-delay operators of Eq.\ \eqref{eq:delays} commute only if the armlengths $L_{ij}$ are constant over times comparable to the longest delays. If this is not the case, ``first-generation'' TDI observables such as $X$ cannot cancel laser frequency noise to the required accuracy. More complex ``second-generation'' TDI observables have been derived that cancel all laser frequency noise residuals proportional to $\dot{L}$; however, for the purpose of computing SNRs and sensitivities, second-generation observables are equivalent to their first-generation analogs \cite{2005PhRvD..72d2003V,2008CQGra..25f5005V}, so we shall consider only the latter in this paper.

The simplest possible space-based interferometric mission involves two ``end-mirror'' spacecraft exchanging laser links with a single ``corner'' spacecraft; in that case, effectively only one observable equivalent to $X$ can be formed. With the classic LISA design, which includes six two-way links between three spacecraft, many different TDI observables are possible, some of which use four one-way measurements, some five, and some all six. However, all these observables can be reconstructed from a set of three\footnote{For classic LISA, the accounting of independent observables goes as follows: start with six one-way inter-spacecraft measurements $y_{ij}$ and three combinations of intra-spacecraft measurements $z_{ij}$ (which must be used in pairs to cancel fiber noise), then ``solve'' for the six lasers that must be canceled out. Thus there remain three independent GW-sensitive measurements, although in the limit of gravitational wavelengths that are large compared to the armlengths, the symmetrized sum of the measurements is insensitive to GWs.} linearly independent observables, such as the unequal-arm Michelsons $X$, $Y$, and $Z$ centered on the three spacecraft \cite{2008CQGra..25f5005V}. Two or three such observables can be used together to increase SNR and improve GW-parameter determination (this insight goes back at least to Cutler \cite{1998PhRvD..57.7089C}).

\subsection{TDI observables and their noise}
In Ref.\ \cite{2008CQGra..25f5005V}, one of us (MV) described how the search for TDI observables (and, specifically, for a basis $\{O_\alpha\}$ of independent observables) can be turned into a problem of linear algebra. This is possible because when the delay operators commute they have a convenient representation in the Fourier domain,
\begin{equation} \fl
\qquad \mathcal{F}[\mathcal{D}_{ij} \cdots \mathcal{D}_{kl} z(t)] =
\Delta_{ij} \cdots \Delta_{kl} \mathcal{F} [z(t)], \;\;
\mathrm{with} \;\; \Delta_{ij} = \rme^{2 \pi \rmi f L_{ij}}
\end{equation}
where $\mathcal{F}$ denotes the Fourier transform.
For instance, in a four-link--only configuration with corner spacecraft labeled by 1, the Fourier-domain dependence of the $y_{ij}$ and $z_{ij}$ measurements on the laser noises is described by
\begin{equation} \fl
\qquad \underbrace{\left(\begin{array}{c}
y_{12} \\ y_{21} \\ y_{13} \\ y_{31} \\ (z_{21} - z_{31}) / 2
\end{array}\right)}_y = \underbrace{
\left(\begin{array}{cccc}
\Delta_{12} & -1 & 0 & 0 \\
-1 & \Delta_{21} & 0 & 0 \\
0 & 0 & \Delta_{13} & -1 \\
0 & 0 & -1 & \Delta_{31} \\
0 & -1 & 0 & 1
\end{array}\right)}_\mathsf{D}
\underbrace{\left(\begin{array}{c}
C_{12} \\ C_{21} \\ C_{13} \\ C_{31}
\end{array}\right)}_C;
\label{eq:Dfour}
\end{equation}
TDI observables correspond to vectors $a$ of time delays such that $a^T y = 0$, i.e., to the solutions of $\mathsf{D}^T a = 0$. For the $\mathsf{D}$ of Eq.\ \eqref{eq:Dfour} the only possible solution, except possibly for a normalization and overall delay, reproduces the $X$ of Eq.\ \eqref{eq:unequalmich}.

The non-laser noise spectrum of TDI observables, a key ingredient to building sensitivity curves, is obtained by writing one more vector equation that describes the $y^\mathrm{op}$ and $y^\mathrm{pm}$ content of each $y_{ij}$ and $z_{ij}$ (no delays are involved); multiplying the equation by the TDI coefficients $a$; building the squared complex norm of the resulting scalar; and taking its noise-realization ensemble average.
Equivalently, we can begin with an equation similar to \eqref{eq:unequalmich}; replace the $y_{ij}$ and $z_{ij}$ with the right-hand sides of Eqs.\ \eqref{eq:ynoisesimple} and \eqref{eq:znoisesimple}; set all the $C_{ij}$ to zero; move to the Fourier domain by switching the $\mathcal{D}_{ij}$ to $\Delta_{ij}$ and all noise quantities to their Fourier transforms; and again build the squared complex norm and take the ensemble average. 
Usually (but not necessarily) one assumes that all the noises of the same kind can be described as uncorrelated Gaussian processes with the same spectral density,
\begin{equation}
\langle y^\mathrm{op}_{ij}(f')^* y^\mathrm{op}_{kl}(f) \rangle =
\frac{1}{2} S^\mathrm{op}(f) \delta(f' - f) \delta_{ij,kl}
\end{equation} 
(and similarly for $y^\mathrm{pm}$) and that all $L_{ij} = L$. For instance, the resulting noise spectrum for $X$ is
\begin{equation}
S_X(f) = 16 \, S^\mathrm{op}(f) \sin^2 x + 16 \, S^\mathrm{pm}(f) (3 + \cos 2x) \sin^2 x,
\label{eq:noiseX1}
\end{equation}
where $x = 2 \pi f L$. For the folded ``syzygy'' \cite{2003PhRvD..68f2001E} linear configuration studied as the 2011 NASA RFI concept \emph{SGO lowest} \cite[see bottom right panel of Fig.\ \ref{fig:tdi}]{nasarfi}, the structure of $X$ does not change, but the differing armlengths (which we describe by setting $L_{23} = L_{31} = L_{32} = L_{13} = L$ and $L_{12} = L_{21} = 2L$), result in different trigonometric multipliers for the noises:
\begin{equation} \fl
S^\mathrm{syzygy}_X(f) = 16 \, S^\mathrm{op}(f) (3 + \cos 2x) \sin^2 x + 
 16 \, S^\mathrm{pm}(f) (5 + 4 \cos 2x + \cos 4x) \sin^2 x,
\end{equation}
where we have also set $S^\mathrm{op}_{12} = S^\mathrm{op}_{21}$ four times larger than all other $S^\mathrm{op}_{ij}$, to account for the greater shot noise due to fewer photons available on the longer link.

\subsection{Bases of uncorrelated TDI observables}
In general, the observables in a basis $\{O_\alpha\}$ have correlated noises, since the same $y^\mathrm{op}_{ij}$ and $y^\mathrm{pm}_{kl}$ can appear in different $O_\alpha$s. (Since the $C_{ij}$ have been canceled out, we do not need to worry about them.) However, by diagonalizing the matrix $S_{\alpha\beta} = \langle O^*_\alpha(f) O_\beta(f) \rangle$ of noise covariances we can find a basis of uncorrelated observables. These are especially useful because the total $\mathrm{SNR}^2$ is then the sum of the individual $\mathrm{SNR}_\alpha^2$.

For the classic-LISA six-link configuration, three solutions to the equivalent of Eq.\ \eqref{eq:Dfour} are given by $X$ and by the $Y$ and $Z$ obtained by cycling indices. The noise covariances turn out to be
\begin{equation}
S_{XY} = S_{YZ} = S_{ZX} = -(8 \, S^\mathrm{op} + 32 \, S^\mathrm{pm}) \cos x \sin^2 x;
\end{equation}
thus a basis of uncorrelated observables that diagonalize the resulting $S_{\alpha\beta}$ is given by the celebrated $A$, $E$, and $T$,
\begin{equation} \fl
\qquad A = (Z - X)/\sqrt{2}, \quad E = (X - 2Y + Z)/\sqrt{6}, \quad T = (X + Y + Z)/\sqrt{3},
\end{equation}
with eigenvalues (i.e., noise spectral densities)
\begin{eqnarray} \fl
\qquad S_{A,E} &= 8\,S^\mathrm{op}(2 + \cos x)\sin^2 x + 16\,S^\mathrm{pm}(3 + 2 \cos x + \cos 2x) \sin^2x, \\ \fl
\qquad S_T &= 16\,S^\mathrm{op}(1 - \cos x)\sin^2 x + 128\,S^\mathrm{pm}\sin^2 x \sin^4 (x/2).
\end{eqnarray}

For the less well-known case of a ``bow tie'' configuration of four spacecraft at the vertices of a square, linked by four two-way links, two along diagonals of length $L$, and two along opposite sides of length $L/\sqrt{2}$ (see top right panel of Fig.\ \ref{fig:tdi}), a correlated basis is given (intuitively) by the four unequal-arm Michelsons $W$, $X$, $Y$, and $Z$ that can be formed around the vertices 1--4 (defined with signs such that $W \rightarrow X \rightarrow Y \rightarrow Z \rightarrow W$ under the cyclical spacecraft relabeling 
$1 \rightarrow 2 \rightarrow 3 \rightarrow 4 \rightarrow 1$). An uncorrelated basis is given by
\begin{equation} \fl
{\textstyle \frac{1}{2}}(W - X + Y - Z), \,
{\textstyle \frac{1}{2}}(W + X - Y + Z), \,
{\textstyle \frac{1}{2}}(W - X - Y + Z), \,
{\textstyle \frac{1}{2}}(W + X + Y + Z),
\end{equation}
with four non-degenerate power spectra, which are somewhat unilluminating functions of $S^\mathrm{op}$, $S^\mathrm{pm}$, $x$, and $x/\sqrt{2}$ that we do not show here.

\section{GW response of space-based interferometers}
\label{sec:signal}

We turn now to the GW response of LISA-like detectors. The ingredients needed for this are the geometry of the constellation (i.e., the individual spacecraft orbits) and the time-dependent GW polarization tensors at the locations of the spacecraft. 
Under the very applicable idealization of GWs as plane waves, and under the provably true assumption that the constant and orbital-Doppler-induced frequency offsets between lasers can be removed once in orbit, the two ingredients come together into the Estabrook--Wahlquist two-pulse response for one-way measurements \cite{1975GReGr...6..439E},
\begin{equation}
y^\mathrm{GW}_{ij}(t) = \frac{1}{2} \frac{\hat{n}_{ij} \cdot
\bigl[\mathsf{h}(p_i,t-L_{ij}) -
\mathsf{h}(p_j,t)\bigr] \cdot  \hat{n}_{ij}}{1 - \hat{n}_{ij} \cdot \hat{k}},
\label{eq:estawahl}
\end{equation}
where we work in SSB coordinates and where:
\begin{itemize}
\item $\hat{n}_{ij}$ is the unit vector from spacecraft $i$ to $j$;
\item $\hat{k}$ is the plane-GW propagation vector;
\item $\mathsf{h}(p_k,t)$ is the TT-gauge GW strain (see, e.g., \cite{Maggiore:1900zz}) at the spacecraft position $p_k$ and time $t$;
\item $L_{ij}$, as before, is the photon time of flight along the arm.
\end{itemize}
Thus, an individual GW pulse appears twice in a time series of each $y_{ij}$: once when it impinges on spacecraft $i$, and a time $L_{ij}$ after it has impinged on spacecraft $j$. The GW response of the $z_{ij}(t)$ is effectively zero, because the intra-spacecraft armlength is so small. The GW response of the TDI observables is then just a sum of delayed $y^\mathrm{GW}(t)$ terms. 

\subsection{Spacecraft orbits}
Since the spacecraft move, the vectors $p_i$ and $\hat{n}_{ij}$ and the armlengths $L_{ij}$ are functions of time: specifically, in Eq.\ \eqref{eq:estawahl} they should appear in full as $p_i(t - L_{ij}(t))$, $p_j(t)$, $\hat{n}_{ij}(t)$, and $L_{ij}(t)$, with the latter two defined implicitly by the light-propagation equation $L_{ij}(t) \hat{n}_{ij}(t) = p_j(t) - p_i(t - L_{ij}(t))$. For many applications, however, it is sufficient to evaluate $\hat{n}_{ij}$,  $p_i$, and $L_{ij}$ at the nominal time of observation $t$.

Equation \eqref{eq:estawahl} can be used for any type of spacecraft orbits, whether specified analytically or numerically. Several of the LISA-like missions concepts under investigation as of late 2011 all adopt variations of the 1 AU heliocentric orbits planned for the three LISA spacecraft. These are slightly eccentric Earth-like orbits that are inclined with respect to the ecliptic plane (by an angle $\simeq \sqrt{3}/2 \times L/\mathrm{AU}$ rad), and that lie on three different planes separated by $120^\circ$, resulting in a quasi-equilateral-triangle spacecraft configuration that is inclined by $60^\circ$ with respect to the ecliptic. While the center of the triangle completes a full revolution around the Sun in the course of a year, the triangle itself (as well as the normal to its plane) rotates through $360^\circ$ in a cartwheeling motion (see Fig.\ \ref{fig:lisa}). The armlengths change slowly, and are always equal to one another to $\sim 1\%$. In \ref{app:motion} we give simple analytic expressions for these orbits; these expressions enforce constant and equal armlengths, and are sufficiently accurate for the computation of sensitivity curves.

\subsection{GW source conventions}
\label{sec:gwsource}
Moving on to GW sources, any plane GW incoming along the propagation vector $\hat{k}$ can be described as
\begin{equation} \fl
\qquad \mathsf{h}(t,x) = \mathsf{h}_\mathrm{SSB}(t - \hat{k} \cdot x) = 
\mathsf{e}_+(\hat{k},\psi) h_+(t - \hat{k} \cdot x) +
\mathsf{e}_\times(\hat{k},\psi) h_\times(t - \hat{k} \cdot x),
\end{equation}
where $\mathsf{h}_\mathrm{SSB}$ is the value of the metric perturbation at the SSB,\footnote{
The time variable $t$ that appears in $h^+(t)$ and $h^\times(t)$ is the time at the origin of the SSB frame, which is related to the time in the GW-source frame by a special- and general-relativistic time dilation; it is however expedient to describe the GW using SSB time, including time-dilation factors such as cosmological redshift in the parametrization of the GW source.} and where $h_+$ and $h_\times$ are the two GW polarizations (defined modulo a polarization rotation encoded by $\psi$), and where $\mathsf{e}_+$ and $\mathsf{e}_\times$ are polarization tensors such that 
$\mathsf{e}_+ : \mathsf{e}_+ = \mathsf{e}_\times : \mathsf{e}_\times = 2$, that $\mathsf{e}_+ : \mathsf{e}_\times = 0$, and that $\mathsf{e}_+ \cdot \hat{k} = \mathsf{e}_\times \cdot \hat{k} = 0$. A possible choice \cite{PhysRevD.70.022003} is to take
\begin{equation} \fl
\qquad \mathsf{e}_+ \equiv \mathsf{O}_1 \cdot \left( \begin{array}{ccc} 1 & 0 & 0 \\ 0 & -1 & 0 \\ 0 & 0 &
0 \end{array} \right) \cdot \mathsf{O}_1^T, \quad
\mathsf{e}_\times \equiv \mathsf{O}_1 \cdot \left( \begin{array}{ccc} 0 & 1 & 0 \\ 1 & 0 &
0 \\ 0 & 0 & 0 \end{array} \right) \cdot \mathsf{O}_1^T,
\label{eq:poltensors}
\end{equation}
with
\begin{equation} \fl
{\sf O}_1 \!=\! \left(\!\!\!
\begin{array}{c@{\,}c@{\,}c}
 \sin\lambda\cos\psi -\cos\lambda\sin\beta\sin\psi  & -\sin\lambda\sin\psi -\cos\lambda\sin\beta\cos\psi & -\cos\lambda\cos\beta\\
-\cos\lambda\cos\psi -\sin\lambda\sin\beta\sin\psi  &  \cos\lambda\sin\psi -\sin\lambda\sin\beta\cos\psi & -\sin\lambda\cos\beta\\
 \cos\beta\sin\psi                              &  \cos\beta\cos\psi                             & -\sin\beta\\
\end{array}
\!\!\right)\!.
\label{eq:polrotation}
\end{equation}
Here $\beta$ and $\lambda$ are the ecliptic latitude and longitude of the GW source, so 
\begin{equation}
\hat{k} = -(\cos\lambda\cos\beta, \sin\lambda\cos\beta, \sin\beta).
\end{equation}
The sky-position parameters $\beta$, $\lambda$, and $\psi$ can be related to the $\theta$, $\phi$, and $\psi$ used in Ref.\ \cite{PhysRevD.67.022001} and in the Mock LISA Data Challenges \cite{2006AIPC..873..619A,2006AIPC..873..625.,2007CQGra..24..551A,2008CQGra..25r4026B,2010CQGra..27h4009B} by setting $\beta = \pi/2 - \theta$, $\lambda = \phi$, and $\psi = -\psi$.

\subsection{The GW response to monochromatic signals}
\label{sec:gwresponse}
Looking ahead to computing the sensitivity to monochromatic GWs, we can specialize Eq.\ \eqref{eq:estawahl} to the GW strain of Eqs.\ \eqref{eq:mono1} and \eqref{eq:mono2}
\begin{eqnarray} \fl \qquad
y^\mathrm{GW}_{ij}(f) = & \frac{
(1 + \cos^2 \iota) \, \hat{n}_{ij} \!\cdot\! \mathsf{e}_+ \!\cdot\! \hat{n}_{ij}
+ \rmi (-2 \cos \iota) \, \hat{n}_{ij} \!\cdot\! \mathsf{e}_\times \!\cdot\! \hat{n}_{ij}
}{2(1 - \hat{n}_{ij} \cdot \hat{k})} \label{eq:ewmono} \\ \fl \qquad
& \times \Bigl(\rme^{2\pi\rmi f_0 (L_{ij} + \hat{k}\cdot p_i)} - \rme^{2\pi\rmi f_0 \hat{k}\cdot p_j}
\Bigr) \frac{A}{2} \rme^{\rmi \phi_0} \delta(f-f_0) \quad \mathrm{for} \, f>0,\nonumber
\end{eqnarray}
where we have used the Fourier-domain representation of time delays. To get the GW response function for a TDI observable, we sum up the individual $y^\mathrm{GW}_{ij}(f)$ for each link, multiplying each by the appropriate time-delay coefficient, and take the square complex amplitude of the resulting expression. For instance, for the $X$ of Eq.\ \eqref{eq:unequalmich} we would write
\begin{eqnarray} \fl
\qquad X^\mathrm{GW}(f) = &
(\Delta_{21} + \Delta_{21}\Delta_{13}\Delta_{31}) y^\mathrm{GW}_{12}(f) +
(1 + \Delta_{12}\Delta_{21}) y^\mathrm{GW}_{21}(f) \label{eq:responsefnX1}
 \\ \fl \qquad & +
(\Delta_{31} + \Delta_{31}\Delta_{12}\Delta_{21}) y^\mathrm{GW}_{13}(f) +
(1 + \Delta_{13}\Delta_{31}) y^\mathrm{GW}_{31}(f). \nonumber
\end{eqnarray}
In practice, the difference of two exponentials on the second row of Eq.\ \eqref{eq:ewmono} can be factored as $\rme^{2\pi\rmi f_0 \hat{k}\cdot p_i} (\rme^{2\pi\rmi f_0 L_{ij}} - \rme^{2\pi\rmi f_0 L_{ij} \hat{k} \cdot \hat{n}_{ij}})$, and $\rme^{2\pi\rmi f_0 \hat{k}\cdot p_i}$ further decomposed by writing $p_i = R + (p_i - R)$, with $R$ the vector to the ``center'' of the spacecraft formation. The factor $\rme^{2\pi \rmi f_0 \hat{k} \cdot R}$ then becomes common to all the $y^\mathrm{GW}_{ij}(f)$ that appear in a TDI observable, and can be dropped (together with $\rme^{\rmi \phi_0}$) when taking the square complex amplitude. As a further simplification, in the limit of equal armlengths all $2\pi f L_{ij} = 2\pi f L \equiv x$, and all $\Delta_{ij} = \rme^{\rmi x}$.

Since the spacecraft are always moving along their orbits, frequency-domain expressions such as Eq.\ \eqref{eq:ewmono} are only accurate for data segments that are short compared to the typical orbital timescale (for LISA-like orbits, a year). However, as long as the length of segments is sufficient to resolve the lowest GW frequency of interest, the response function can be obtained easily by averaging the individual ``instantaneous'' responses. This approximation neglects the Doppler modulation of individual monochromatic sources, which changes results only marginally by smoothing out
signals across nearby frequencies.

For the spacecraft orbits of \ref{app:motion}, which are described by a simple geometric transformation, we can take the alternative, expedient approach of deriving the apparent motion of a GW source in the observatory frame [see the end of \ref{app:motion}], and then integrate the stationary-detector sensitivity along that path [see the discussion around Eq.\ \eqref{eq:movingeffstrain1}].

\section{Sensitivity to isotropic and Galactic distributions of quasi-monochromatic sources}
\label{sec:sensitivity}

We now draw together the ingredients introduced in Secs.\ \ref{sec:noise} and \ref{sec:signal} to calculate the averaged sensitivity of a LISA-like detector to quasi-monochromatic sources, and to characterize its variation over sky positions and orientations. Because the detector response to GWs is directional, these quantities are only defined with respect to a distribution of sources in the sky (i.e., a \emph{population}): here we consider the simplest, isotropic distribution, as well as a realistic Galactic-disk distribution. The latter analysis is relevant, for instance, to the sensitivity of LISA-like observatories to verification binaries \cite{2006CQGra..23S.809S}. Since verification binaries have known positions, it is possible to compute their exact $\mathrm{SNR}_\mathrm{opt}$ (modulo unknown inclinations), but the averaged sensitivity and its ``error bar'' is still useful for quick calculations, or for yet-to-be-discovered verification systems.

Traditionally, the isotropic sky-averaged LISA sensitivity has been computed for a stationary detector configuration, reasoning that orbital motion can be neglected because, in the LISA ``frame,'' an isotropic source distribution remains invariant as the detector's orientation changes along the year, so the sky-averaged sensitivities of stationary and moving detectors would coincide. In this section, we compute sensitivities in both cases, and show that the argument is indeed correct for the averaged sensitivity, but that the resulting sensitivity distributions are rather different. For instance, the 5\%-quantile sensitivity is $\sim 5$ times worse than the average sensitivity for a stationary detector, but only $\sim 2$ times worse for a moving detector, whose motion effectively smears away the ``blind spots'' of the stationary detector.

This section is organized as follows: in Sec.\ \ref{sec:sensproc} we describe the details of our calculation, including our assumptions for the instrument noises, for the residual confusion noise from the Galactic white-dwarf background, and for our Galactic-disk population. In Sec.\ \ref{sec:sensresults} we plot the resulting distributions of sensitivities, discuss their features, and provide frequency-dependent fits.

\subsection{Computing the sensitivity}
\label{sec:sensproc}
\begin{table}
\footnotesize \flushright
\begin{tabular}{@{}l@{\,\,}l@{}}
\hline \hline
\multicolumn{1}{c}{$(S^\mathrm{op})^{1/2}$, physical} & \multicolumn{1}{c}{$(S^\mathrm{pm})^{1/2}$, physical} \\
\multicolumn{1}{c}{$S^\mathrm{op}$, fractional frequency} & \multicolumn{1}{c}{$S^\mathrm{pm}$, fractional frequency} \\ \hline \hline
\multicolumn{2}{c}{\emph{Mock LISA data challenges}} \\
$20 \times 10^{-12} \, \mathrm{m} \, \mathrm{Hz}^{-1/2}$
& $3 \times 10^{-15} [1 + (\frac{f_\mathrm{Hz}}{10^{-4}})^{-2}]^{1/2} \, \frac{\mathrm{m} \, \mathrm{s}^{-2}}{\mathrm{Hz}^{1/2}}$ \\
$1.8 \times 10^{-37} f_\mathrm{Hz}^2 \, \mathrm{Hz}^{-1}$
& $2.5 \times 10^{-48} [1 + (\frac{f_\mathrm{Hz}}{10^{-4}})^{-2}] \, f_\mathrm{Hz}^{-2} \, \mathrm{Hz}^{-1}$ \\
\hline
\multicolumn{2}{c}{\emph{LISA science requirements} (2007--2011)} \\
$18 \times 10^{-12} [1 + (\frac{f_\mathrm{Hz}}{2 \times 10^{-3}})^{-4}]^{1/2} \, \mathrm{m} \, \mathrm{Hz}^{-1/2}$
& $3 \times 10^{-15} [1 + (\frac{f_\mathrm{Hz}}{10^{-4}})^{-1}]^{1/2} [1 + (\frac{f_\mathrm{Hz}}{8 \times 10^{-3}})^4]^{1/2} \, \frac{\mathrm{m} \, \mathrm{s}^{-2}}{\mathrm{Hz}^{1/2}}$ \\ 
$1.42 \times 10^{-37} [1 + (\frac{f_\mathrm{Hz}}{2 \times 10^{-3}})^{-4}] \, f_\mathrm{Hz}^2 \, \mathrm{Hz}^{-1}$
& $2.53 \times 10^{-48} [1 + (\frac{f_\mathrm{Hz}}{10^{-4}})^{-1}] [1 + (\frac{f_\mathrm{Hz}}{8 \times 10^{-3}})^4] \, f_\mathrm{Hz}^{-2} \, \mathrm{Hz}^{-1}$ \\ \hline
\multicolumn{2}{c}{\emph{ESA NGO} (2011, final)} \\
$(5.25 \times 10^{-23} + 6.28 \times 10^{-23})^{1/2}\, \mathrm{m} \, \mathrm{Hz}^{-1/2}$
& $4.6 \times 10^{-15} [1 + (\frac{f_\mathrm{Hz}}{10^{-4}})^{-1}]^{1/2} \, \frac{\mathrm{m} \, \mathrm{s}^{-2}}{\mathrm{Hz}^{1/2}}$ \\ 
$(2.31 \times 10^{-38} + 2.76 \times 10^{-38}) f_\mathrm{Hz}^2 \, \mathrm{Hz}^{-1}$ 
& $6 \times 10^{-48} [1 + (\frac{f_\mathrm{Hz}}{10^{-4}})^{-1}] f_\mathrm{Hz}^{-2} \, \mathrm{Hz}^{-1}$ \\ \hline \hline
\end{tabular}
\caption{Optical-path and proof-mass noise components for three models of LISA-like GW observatories. Noises are given in physical units and in the fractional-frequency fluctuations units used in this paper. Note that for ESA NGO, the reference armlength is 1 Mkm rather than LISA's 5 Mkm. The two $S^\mathrm{op}$ components given for ESA NGO correspond to shot noise (which scales with armlength, laser power, and telescope size) and to other fixed optical noises.\label{tab:noises}}
\end{table}

We now proceed to compute the LISA-like sensitivity to monochromatic-binary GW signals for a single TDI observable $X$. We consider two detector geometries: i) a stationary equilateral-triangle constellation with 5 Mkm arms, lying in the ecliptic plane and centered at the Sun; ii) a rotating constellation in solar orbit, with the LISA-like spacecraft trajectories described in \ref{app:motion}.

For both cases we randomly select $10^5$ source positions in the sky, distributed either isotropically (uniformly in $\sin \beta$ and $\lambda$), or across a thin-disk Galactic model of density
\begin{equation}
\frac{dN}{\mathrm{d}r\,\mathrm{d}z}
\propto r \exp(-r/2.5 \, \mathrm{kpc}) \, \mathrm{sech}(z/300 \, \mathrm{pc})
\label{eq:galacticdisk}
\end{equation}
\cite{1988A&A...192..117V}, where $r$ and $z$ are the Galactic cylindrical radius and height (they can be converted to ecliptic $\beta$ and $\lambda$ following, e.g., \cite{roth2009handbook}). For each source position we also randomly select the inclination $\iota$ and polarization angle $\psi$ from isotropic distributions.

We use the instrument noise of Eq.\ \eqref{eq:noiseX1}, with the assumptions on optical-path and proof-mass noises given in the 2011 LISA requirements \cite{lisasens}, and spelled out in Table \ref{tab:noises}, where they are compared with the parameters used for the Mock LISA Data Challenges \cite{2006AIPC..873..619A,2006AIPC..873..625.,2007CQGra..24..551A,2008CQGra..25r4026B,2010CQGra..27h4009B} and for the ESA 2011 ``NGO'' study \cite{esastudy,amaldiproc}.

The instrument, however, is not the only source of noise that affects the sensitivity. LISA-like detectors with sufficiently long baselines and low instrument noise will be subject to an unresolvable GW confusion foreground from white-dwarf binaries in the Galaxy (see \cite{nvpn2012} for a recent treatment and extensive references). In this paper we adopt the confusion-foreground model developed in \cite{nvpn2012},
\begin{equation} \fl
S_X^\mathrm{conf} (f) =
\left[ 1.4 \times 10^{-45} (f/\mathrm{Hz})^{-8/3} \, \mathrm{Hz}^{-1} \right] \times
\tanh^\alpha \! \left( \frac{\beta}{2} \frac{\mathrm{d}N}{\mathrm{d}f} \right) \times
\left[ \frac{3}{20} \frac{ 16 \, x^2 \sin^2 x}{(1 + 0.6 \, x^2)} \right]
\label{eq:confusionnoise1}
\end{equation}
(here expressed as fractional frequency fluctuations for the TDI observable $X$),
where $\alpha \simeq 1$, $\beta \simeq 1$/yr, $\mathrm{d}N/\mathrm{d}f = 5 \times 10^{-3} (f/\mathrm{Hz})^{-11/3} \, \mathrm{Hz}^{-1}$, and $x = 2 \pi f L$.
The first term in Eq.\ \eqref{eq:confusionnoise1} is a fit of the GW-strain power spectral density of unresolved Galactic-binary sources, in the mid-frequency regime where systems are evolving because of GW emission alone; the second term is a fit of the transition to the higher frequencies (above a few mHz) where binary signals are sufficiently spaced apart in frequency domain to be resolved and subtracted individually; the third term is a sky-averaged, approximated transfer function from equivalent strain to $X$. The total $X$ noise is then given by $S_X(f) + S^\mathrm{conf}_X(f)$, and is comparable to (but measurably different from) the often-seen expression proposed by Barack and Cutler \cite{BCu04B}.

When we compute moving-detector sensitivities, we include the strong seasonal variation in the amplitude of the confusion foreground caused by the evolving orientation of the LISA-like formation with respect to the Galaxy. Following Ref.\ \cite{nvpn2012}, and consistently with the orbits of \ref{app:motion}, we set
\begin{eqnarray} \fl
\quad \quad \quad S_X^\mathrm{conf}(f,t) = S_X^\mathrm{conf}(f) \times \bigl[1 &-& 0.21 \cos(2 \pi t/\mathrm{yr}) - 0.45 \cos(4 \pi t/\mathrm{yr}) \\ &-& 0.05 \sin(2 \pi t/\mathrm{yr}) + 0.01 \sin(4 \pi t/\mathrm{yr}) \bigr]. \nonumber
\end{eqnarray}

For each source geometry (sky position, inclination, and polarization, denoted collectively by $\Omega$), we compute a \emph{full} sensitivity curve from $f = 10^{-4}$ to $10^{-1}$ Hz. Specifically, we insert the binary quadrupole-emission polarization tensor [the first bracket of Eq.\ \eqref{eq:mono2}] into the frequency-dependent GW response $X^\mathrm{GW}$ of Eqs.\ \eqref{eq:ewmono} and \eqref{eq:responsefnX1}, using the $\mathsf{e}_+$ and $\mathsf{e}_\times$ of Eqs.\ \eqref{eq:poltensors} and \eqref{eq:polrotation}. The effective strain noise (i.e., the TDI noise over the $\Omega$-dependent squared TDI response function) is then
\begin{equation}
S_{\rm eff}(f; \Omega) = \frac{ S_X (f) + S_X^{\rm conf}(f) }{ |X^{\rm GW} (f; \Omega) |^2 },
\end{equation}
and the corresponding $\Omega$-dependent monochromatic-signal sensitivity is
\begin{equation}
{\rm sensitivity} (f_0; \Omega) = {\rm SNR}_{\rm thr} \frac{ \sqrt{ S_{\rm eff} (f_0; \Omega) } }{ \sqrt{T} }.
\label{eq:monosensstationary1}
\end{equation}
In the next section, we study the distribution of this sensitivity function over our sample of geometries $\Omega$. We define the average sensitivity function from the inverse-rms averaged effective noise $S_\mathrm{eff}^\mathrm{rms}(f) = \langle S_\mathrm{eff}(f;\Omega)^{-1} \rangle_\Omega^{-1}$.
We always set ${\rm SNR}_{\rm thr} = 5$ and $T = 1$ yr.

As mentioned at the end of Sec.\ \ref{sec:gwresponse} and in \ref{app:motion}, we handle the moving-detector case by imposing an apparent motion $\Omega'(t;\Omega)$ on the sources, and integrate the resulting instantaneous effective strain noise,
\begin{equation}
S^\mathrm{moving}_\mathrm{eff} (f;\Omega) = \int_0 ^T dt \, \frac{ S_X (f) + S^\mathrm{conf}_X(f,t)}{ \left| X^{\rm GW}(f;\Omega'(t;\Omega)) \right|^2 }
\label{eq:movingeffstrain1}
\end{equation}
In practice, we compute this integral as a discrete sum of terms evaluated at $N_{\rm pos}$ positions along the year-long LISA-like orbits. We find that $N_{\rm pos} = 26$ provides a sufficiently accurate approximation, with a typical error $\ll 1\%$ (a larger $N_\mathrm{pos}$ will be required at frequencies higher than we consider here).
\begin{figure}
\includegraphics[width=\textwidth]{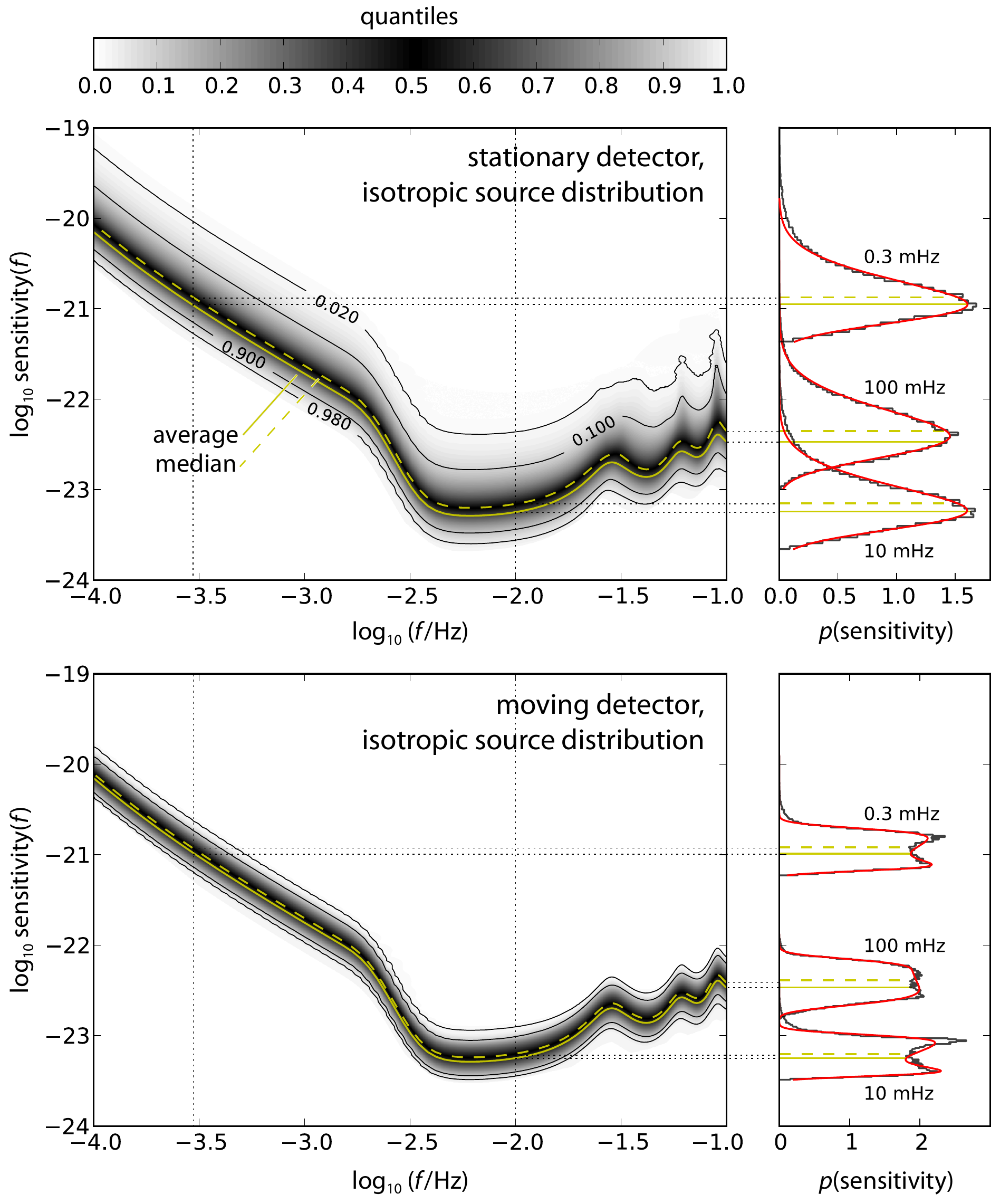}
\caption{
\textbf{Top panel}: Distribution of sensitivities for a stationary LISA-like detector, over a sample of $10^5$ source geometries selected isotropically across the sky.
Gray levels map the quantiles of the distribution (i.e., the fraction of the total sample for which sensitivity is better than plotted); the four black curves correspond to $2\%$, $10\%$, $90\%$, and $98\%$ quantiles. The inverse-rms averaged and median sensitivities are plotted by the solid and dashed yellow lines, respectively. The panel on the right shows histograms of sensitivities for representative frequencies of $0.3$, $10$, and $100$ mHz; these are fit very well by the skew-normal distributions of Eq.\ \eqref{eq:skew1}, plotted by the red curves. 
\textbf{Bottom panel}: Distributions of sensitivities for a moving LISA-like detector. The histograms in the right panel are fit well by the modified, double-peaked skew-normal distributions of Eq.\ \eqref{eq:skew2}.}
\label{fig:sensitivities1}
\end{figure}

\subsection{Sensitivity curves: distributions and averages}
\label{sec:sensresults}

Figure \ref{fig:sensitivities1} shows the distribution and average of the sensitivity, computed over the isotropic $\Omega$ sample as discussed in the previous section, for the stationary (top panel) and moving (bottom panel) detector configurations. At any frequency, the vertical gray-level profile encodes the \emph{quantiles} of the sensitivity distribution: for instance, the sensitivity would lie above the curve labeled ``0.020'' (or below the curve labeled ``0.980'') for 2\% of source geometries. The continuous and dashed light-colored lines mark the (inverse-rms) average and median sensitivities, while the histograms on the right show the distributions at three representative frequencies, for which we also give numerical values in Table \ref{tbl:table1}.
\begin{table}
\flushright
\begin{tabular}{rl@{\;/\;}ll@{\;/\;}ll@{\;/\;}l}
	\hline \hline
	~ & \multicolumn{6}{c}{sensitivity, \emph{stationary detector} (absolute/relative to average)} \\
	& \multicolumn{2}{c}{$0.3$ mHz} & \multicolumn{2}{c}{$10$ mHz} & \multicolumn{2}{c}{$100$ mHz}  \\
	\hline
	$5\%$ & $5.26 \times 10^{-21}$ & 4.88 & $2.58 \times 10^{-23}$ & 4.65 & $1.60 \times 10^{-22}$ & 4.71 \\
	$10\%$ & $3.52 \times 10^{-21}$ & 3.26 & $1.78 \times 10^{-23}$ & 3.20 & $1.14 \times 10^{-22}$ & 3.37 \\
	average & $1.08 \times 10^{-21}$ & 1 & $5.54 \times 10^{-24}$ & 1 & $3.40 \times 10^{-23}$ & 1 \\
	median & $1.32 \times 10^{-21}$ & 1.22 & $6.82 \times 10^{-24}$ & 1.23 & $4.46 \times 10^{-23}$ & 1.31 \\
	$90\%$ & $6.96 \times 10^{-22}$ & 0.65 & $3.57 \times 10^{-24}$ & 0.64 & $2.18 \times 10^{-23}$ & 0.64 \\
	$95\%$ & $6.01 \times 10^{-22}$ & 0.56 & $3.08 \times 10^{-24}$ & 0.56 & $1.84 \times 10^{-23}$ & 0.54 \\
	\hline\hline
	~ & \multicolumn{6}{c}{sensitivity, \emph{moving detector} (absolute/relative to average)} \\
	& \multicolumn{2}{c}{$0.3$ mHz} & \multicolumn{2}{c}{$10$ mHz} & \multicolumn{2}{c}{$100$ mHz}  \\
	\hline
	$5\%$ & $2.00 \times 10^{-21}$ & 1.93 & $1.06 \times 10^{-23}$ & 1.91 & $6.84 \times 10^{-23}$ & 2.01 \\
	$10\%$ & $1.82 \times 10^{-21}$ & 1.76 & $9.56 \times 10^{-24}$ & 1.72 & $6.29 \times 10^{-23}$ & 1.85  \\
	average & $1.04 \times 10^{-21}$ & 1 & $5.55 \times 10^{-24}$ & 1 & $3.40 \times 10^{-23}$ & 1 \\
	median & $1.16 \times 10^{-21}$ & 1.12 & $6.21 \times 10^{-24}$ & 1.12 & $3.87 \times 10^{-23}$ & 1.14 \\
	$90\%$ & $7.29 \times 10^{-22}$ & 0.70 & $3.90 \times 10^{-24}$ & 0.70 & $2.39 \times 10^{-23}$ & 0.70 \\
	$95\%$ & $6.85 \times 10^{-22}$ & 0.66 & $3.68 \times 10^{-24}$ & 0.66 & $2.16 \times 10^{-23}$ & 0.64 \\
	\hline \hline
\end{tabular}
\caption{Sensitivities of an isotropic population of monochromatic sources for stationary and moving LISA configurations listed as (absolute sensitivity)/(sensitivity relative to average) for the $5\%$, $10\%$, $90\%$, $95\%$ quantiles, the average, and the median sensitivities for source frequencies of $0.3$, $10$, and $100$ mHz.\label{tbl:table1}}
\end{table}

For the stationary detector, the distributions have long tails: the extremal values in our $\Omega$ sample can be as much as three orders of magnitude worse than the average, corresponding to ``blind spots'' in the sky, caused by unfavorable orientations of the spacecraft constellation with respect to the polarization of the incoming GWs. However, these tails are thin, as shown by the position of the 2\% and 98\% quantile curves. Over the entire frequency range, the average sensitivity corresponds approximately ($\pm 5\%$) to the 67\% quantile curve---thus, about $33\%$ of all sampled source geometries have better (lower) sensitivity.

For the moving detector, the distributions are tighter, with ``error bars'' a factor $\sim 2$ smaller, and almost no tails: since the orientation of the detector with respect to any source changes across the year, any blind spot is smeared out. The average sensitivity corresponds approximately to the $60\%$ quantile curve. As expected, the stationary- and moving-detector average sensitivities coincide very accurately. In \ref{app:snr2fit} we provide a fit that is therefore valid for both.

\begin{figure}
\includegraphics[width=\textwidth]{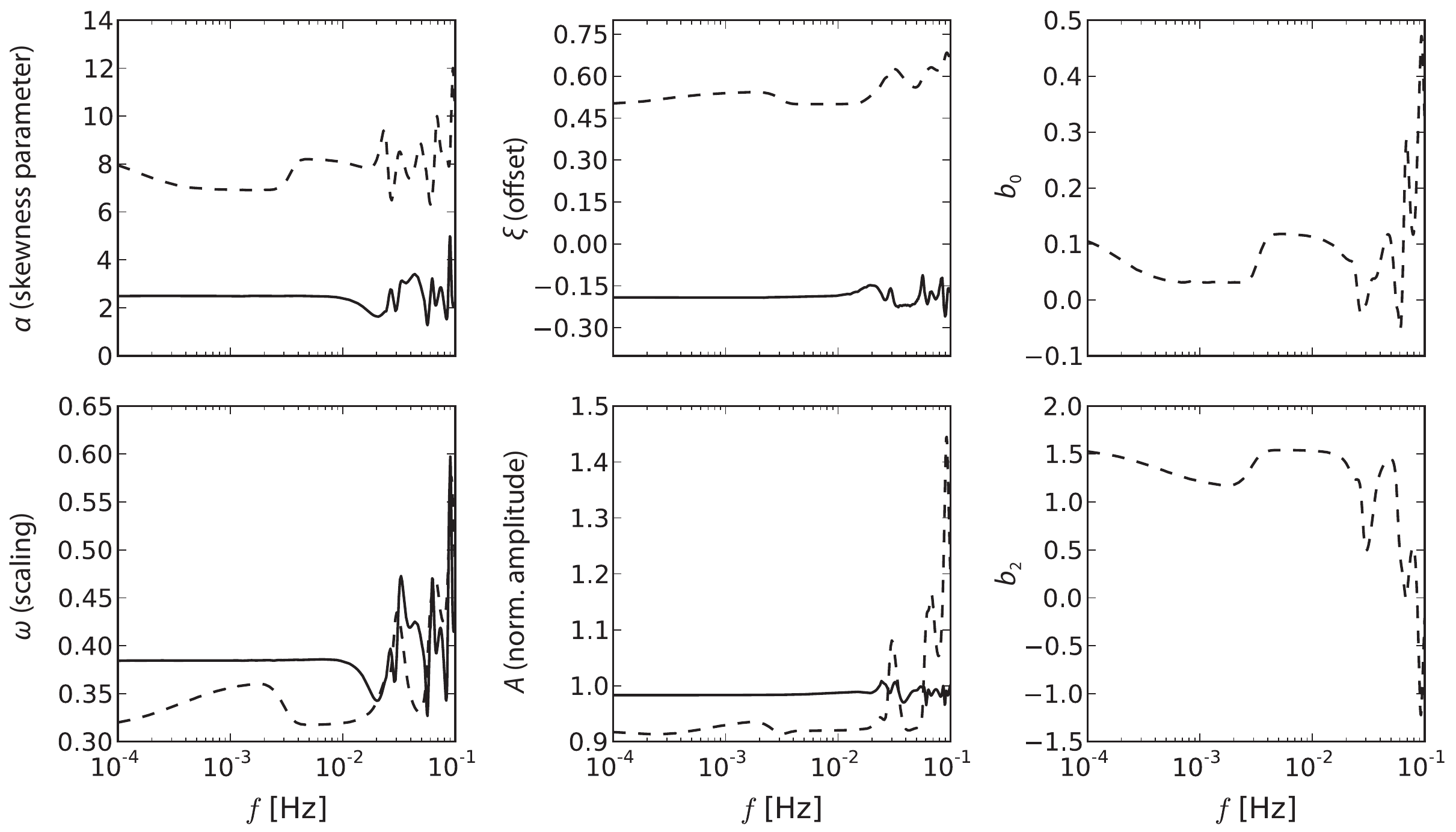}
\caption{Skew-normal and modified skew-normal fit parameters $\alpha$, $\omega$, $\xi$, $A$, $b_0$, and $b_2$, as functions of source frequency $f$, for stationary (solid) and moving (dashed) LISA-like detectors and an isotropic monochromatic-source population.\label{fig:histogramfits1}}
\end{figure}

We find that the stationary-detector sensitivity distribution is fit well, at any frequency in the range shown, by the skew-normal distribution
\begin{equation}
{\rm Skew} (z ; A, \xi , \alpha , \omega) = {} \frac{ A}{ \omega \sqrt{2\pi} } \exp \left( - \frac{ z^2 }{ 2 } \right) \left[ 1 + {\rm erf} \left( \frac{ \alpha z }{ \sqrt{2}  } \right) \right],
\label{eq:skew1}
\end{equation}
where $z \equiv (y - \xi) / \omega$ and where
\begin{equation}
y = \log_{10} \frac{ {\rm sensitivity} (f; \Omega) }{ \langle {\rm sensitivity} (f; \Omega) \rangle_{\mathrm{inverse\,rms}} }.
\label{eq:y1}
\end{equation}
The parameters of Eq.~(\ref{eq:skew1}) specify the shape ($\alpha$), scale ($\omega$), center ($\xi$), and normalization ($A$) of the skew normal. These are all functions of frequency and are shown by the solid curves in Fig.\ \ref{fig:histogramfits1}. For $[ 0.1, 10]$ mHz, the parameters are approximately constant, with $\alpha \approx 2.5$, $\omega \approx 0.38$, $\xi \approx -0.19$, and $A \approx 0.98$.

The moving-detector sensitivity distribution is qualitatively different than the stationary-detector case, and has a somewhat peculiar double-peaked, nearly symmetrical shape.
Surprisingly, it is fit rather well by a slight variation of (\ref{eq:skew1}) that replaces the argument of the exponential with a fourth-order polynomial,
\begin{equation} \fl
\quad {\rm Skew} (z; A, \xi , \alpha,  \omega, b_0, b_2 ) = \frac{A}{\omega \sqrt{ 2\pi} } \exp \bigg( - \frac{ ( z^2 - b_2 z - b_0 ) ^2 }{ 2 } \bigg) \bigg[ 1 + {\rm erf} \bigg( \frac{ \alpha z }{ \sqrt{2} } \bigg) \bigg],
\label{eq:skew2}
\end{equation}
with the two additional structure parameters $b_0$ and $b_2$. All six moving-detector fit parameters are shown as functions of frequency by the dashed curves of Fig.\ \ref{fig:histogramfits1}. Unlike the stationary-detector case, the parameters are sensitive to the presence of confusion noise from the Galactic white-dwarf foreground, as shown by the characteristic bumps below a few mHz.

\begin{figure}
\includegraphics[width=\textwidth]{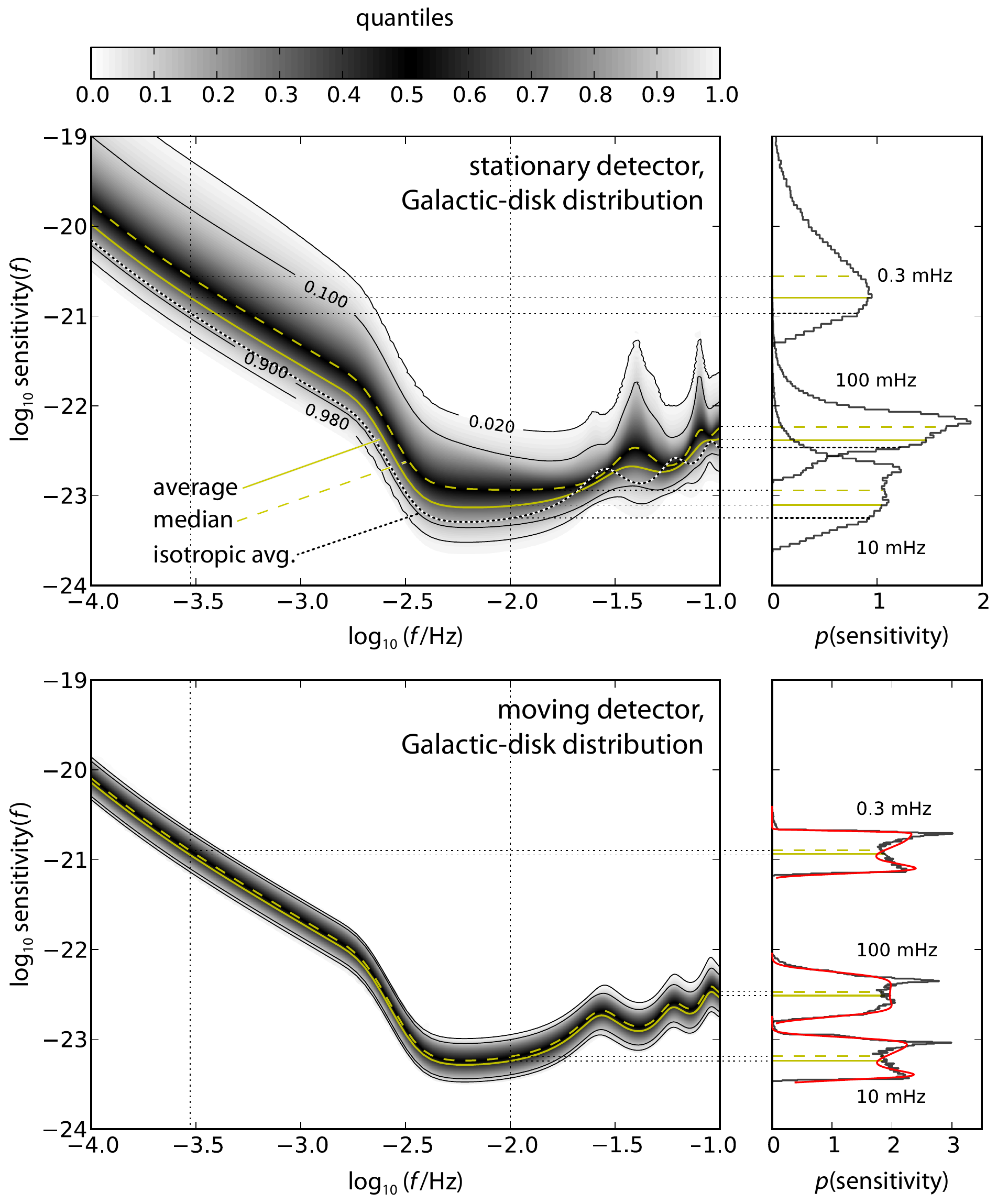}
\caption{
\textbf{Top panel}:  Distribution of sensitivities for a stationary LISA-like detector, over a sample of $10^5$ geometries for sources distributed across the Galactic disk of Eq.\ \eqref{eq:galacticdisk}. All curves are to be interpreted as in Fig.\ \ref{fig:sensitivities1}; the additional dotted black curve plots the average sensitivity for the isotropic source distribution (the yellow solid curve of Fig.\ \ref{fig:sensitivities1}).
\textbf{Bottom panel}: Distribution of sensitivities for a moving LISA-like detector.}
\label{fig:sensitivities2}
\end{figure}

Last, we consider a sample of source geometries distributed according to the Galactic-disk population of Eq.\ \eqref{eq:galacticdisk}. As mentioned above, this analysis is relevant to the sensitivity of LISA-like detectors to verification binaries, which are in fact rather good realizations of monochromatic GW sources. Again, Fig.\ \ref{fig:sensitivities2} shows the distribution and average of the sensitivity, for the stationary (top panel) and moving detector (bottom panel) configurations. For the stationary case, we see that the deviations from the average sensitivity can be much larger than for the isotropic population of Fig.\ \ref{fig:sensitivities1}. The average sensitivities themselves are rather different: compare the solid yellow line with the dotted black line. For a moving detector, sensitivities are concentrated much more tightly around the average, and again exhibit double peaks. Furthermore, the average sensitivity is essentially indistinguishable from the isotropic case.

\begin{figure}
\includegraphics[width=\textwidth]{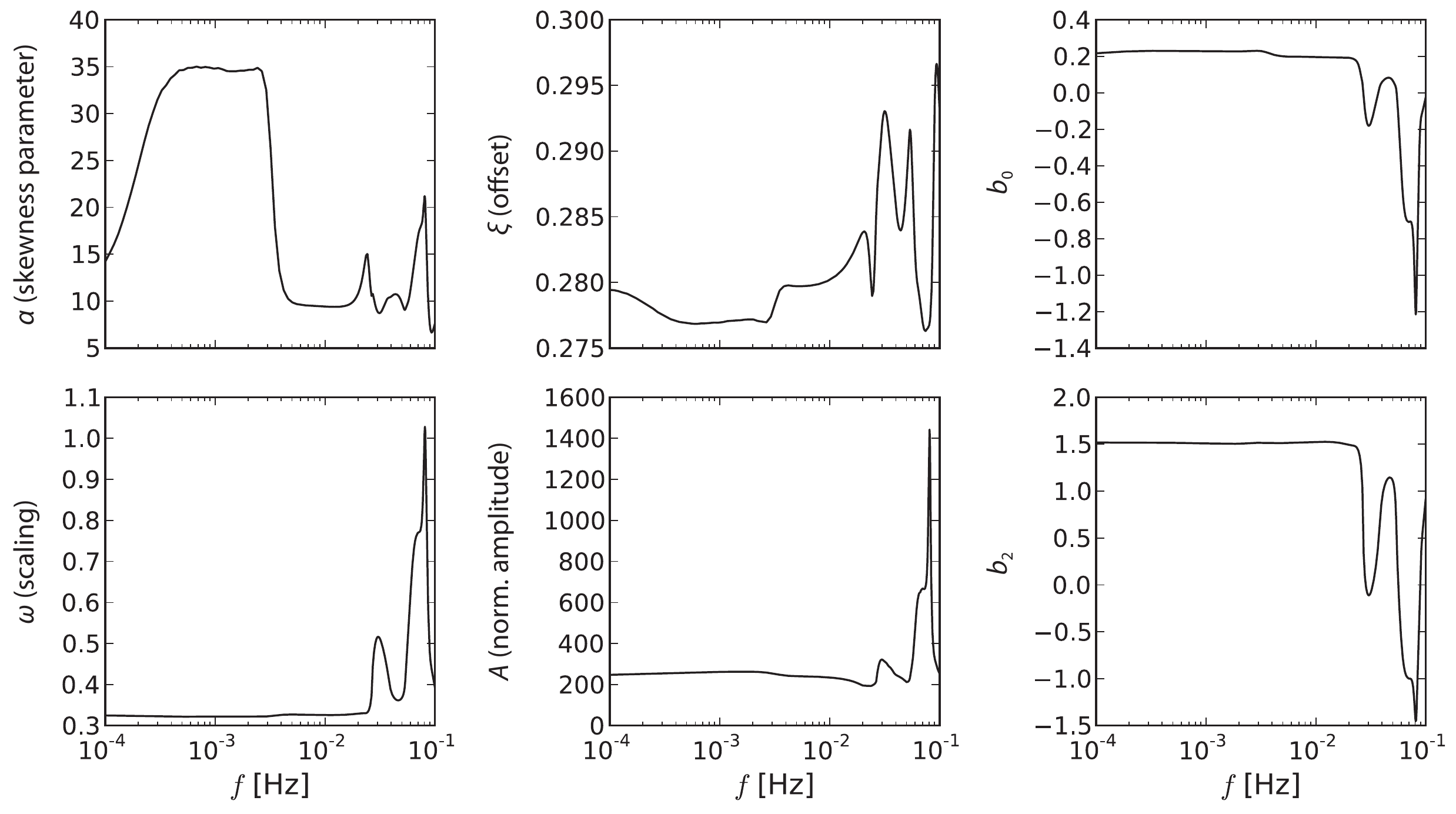}
\caption{Modified skew-normal fit parameters $\alpha$, $\omega$, $\xi$, $A$, $b_0$, and $b_2$, as functions of source frequency $f$, for moving LISA-like detectors and a population of monochromatic sources from the Galactic disk.}
\label{fig:histogramfits2}
\end{figure}

We were unable to find high-accuracy fits to these distributions. The skew normal of Eq.\ \eqref{eq:skew1} works well for a stationary detector at frequencies less than 10 or 20 mHz, but fails dramatically at higher frequencies. The modified skew normal of Eq.\ \eqref{eq:skew2} works reasonably well for a moving detector at all frequencies, although it consistently fails to resolve accurately the sharp peaks of the histograms seen in the bottom-panel of Fig.\ \ref{fig:sensitivities2}. Nevertheless, we do recommend that Eq.\ \eqref{eq:skew2} be used to provide error bars for verification-binary sensitivities. In practice, this can be attained by looking up the values of the modified skew-normal parameters at the desired source frequency in Fig.\ \ref{fig:histogramfits2} and substituting those values into Eq.\ \eqref{eq:skew2}. This yields a function of $z = (y- \xi) /\omega$ where $y$ is given in Eq.\ \eqref{eq:y1} and the inverse-rms sensitivity at the desired source frequency can be found for the appropriate spacecraft configuration from Figs.\ \ref{fig:sensitivities1} and \ref{fig:sensitivities2} or from our fitting functions in Eq.\ \eqref{improvedfit1}. If using the latter one must add the confusion noise to the function defined in Eq.\ \eqref{improvedfit1}. The resulting function of the sensitivity yields the approximate distribution at the given frequency.

\section{Summary and applications}
\label{sec:conclusions}

In this paper we present a straightforward end-to-end recipe to compute the non-sky-averaged sensitivity of interferometric space-based GW detectors. The recipe incorporates the motion of the spacecraft and the seasonal variations in the partially subtracted confusion foreground from Galactic binaries. By evaluating the sensitivity over a sample of source geometries, we are able to provide a stringent statistical interpretation for previously unqualified statements about sky-averaged SNRs. As an example, we consider the ``classic LISA'' sensitivity to monochromatic GW sources that are distributed isotropically or according to a Galactic-disk population.

For an isotropic source population, we confirm that the inverse-rms sky-averaged sensitivity is the same whether or not the motion of the detector is taken into account. However, the stationary-detector calculation overestimates the variance of the sensitivity distribution (see Fig.\ \ref{fig:sensitivities1} and Table \ref{tbl:table1}),
because with orbital motion the ``blind spots'' of the detector trace a path across the sky, and are averaged out over a year.
For a Galactic-disk population, the stationary- and moving-detector average sensitivities are rather different (see Fig.\ \ref{fig:sensitivities2}); indeed, the former will depend strongly on the arbitrary orientation of the stationary detector. However, the moving-detector curve coincides closely with the isotropic average sensitivity. Thus, the standard approximate sensitivity expression described in \ref{app:snr2fit} (where it is also improved with the correct high-frequency behavior) can be used for both populations. The modified skew-normal fit developed in Sec.\ \ref{sec:sensresults} can be used to describe the statistical distribution of sensitivities as a function of frequency, thus quantifying the errors in future studies of GW-source detectability. Approximate fit parameters can be looked up in Fig.\ \ref{fig:histogramfits1} or Fig.\ \ref{fig:histogramfits2} and substituted in Eq.\ \eqref{eq:skew2}.

Our recipe can be easily adapted, and our analysis repeated, for different detector geometries and orbits; indeed, we have done so for the mission concepts submitted to the NASA RFI. For eLISA/NGO, we have confirmed that the distributions of sensitivities are fit well by Eqs.\ \ref{eq:skew1} and \ref{eq:skew2}, albeit with different fitting parameters. Our model of NGO included a constellation with the orbits of \ref{app:motion} but with 1-Mkm armlengths, the instrument noises listed in Table \ref{tab:noises}, plus confusion noise given by Eq.\ \eqref{eq:confusionnoise1} with $\alpha \simeq 0.7$, $\beta \simeq 1.2/\mathrm{yr}$ \cite{nvpn2012}.

Sensitivity curves can also be obtained for different types of sources, such as binary black-hole inspirals. However, since these signals are broadband, it is more appropriate to derive the horizon distance (e.g., as a function of total or chirp mass) rather than the sensitivity as a function of frequency. Furthermore, the relative timing of the inspiral and detector orbits also enters the calculation, and needs to be included in the statistical sample. To facilitate such studies, we are providing a \textsl{Mathematica} notebook that implements our calculation, which can be easily modified for different detectors and sources, which includes fitting parameters for ``our'' LISA and NGO, and which can be found in the CQG supplementary materials for this paper, and at \url{http://www.vallis.org/publications/sensitivity}.

\ack MV is grateful to J.\ Armstrong, F.\ Estabrook, and M.\ Tinto for teaching him about TDI and sensitivity calculations. This work was carried out at the Jet Propulsion Laboratory, California Institute of Technology, under contract with the National Aeronautics and Space Administration. MV was supported by the LISA Mission Science Office and by the JPL RTD program. CG was supported by an appointment to the NASA Postdoctoral Program at the JPL administered by Oak Ridge Associated Universities through a contract with NASA. Copyright 2012 California Institute of Technology.

\appendix

\section{A simple geometric model for the motion of LISA-like GW interferometers}
\label{app:motion}

\begin{figure}
\includegraphics[width=\textwidth]{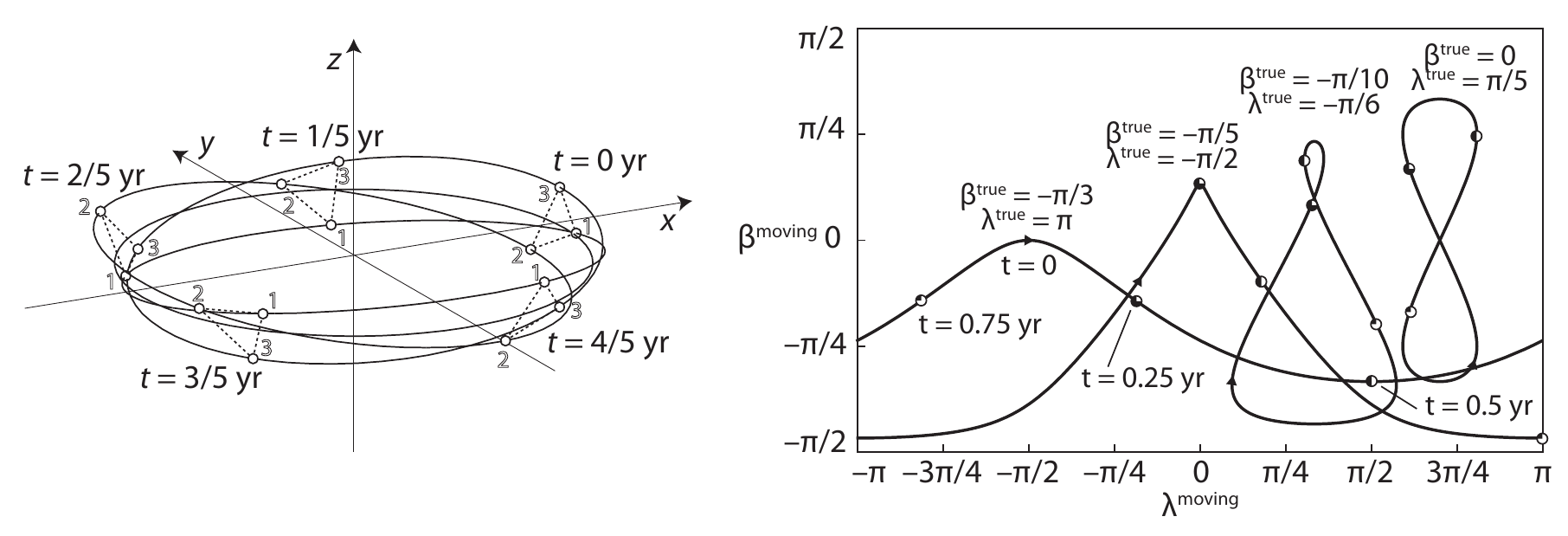}
\caption{\textbf{Left}: Orbital motion of a LISA-like GW interferometer, as described by the simple geometric model of \ref{app:motion}, shown in an ecliptic SSB coordinate system.
\textbf{Right}: Apparent motion in the rotating detector frame of sources at ecliptic sky coordinates $(\beta^\mathrm{true},\lambda^\mathrm{true})$.
\label{fig:lisa}}
\end{figure}

We begin by placing the spacecraft along the vertices of an equilateral triangle, working in a moving, rotating coordinate frame where the spacecraft are at rest \cite{PhysRevD.70.022003}:
\begin{equation} \fl
\qquad p^\mathrm{rest}_i = (L/\sqrt{3}) ( -\cos2\sigma_i , \sin2\sigma_i, 0)
\;\mathrm{with} \; \sigma_i = 3\pi/2 - 2(i-1)\pi/3,
\end{equation} 
where $L$ is the common armlength. Next, we switch to an inertial, ecliptic coordinate system, with origin at the Solar-system barycenter (SSB). In this frame, the center $R$ of the LISA-like formation moves in the $x$--$y$ plane on an SSB-centric circular orbit with radius $1\,\mathrm{AU}$, and the three spacecraft rotate around $R$:
\begin{equation} \fl
\qquad p_i(t) = R + \mathsf{O}_2 \cdot p^\mathrm{rest}_i,
\label{eq:poslisa}
\end{equation}
where $R = (1 \, \mathrm{AU}) \times (\cos\eta, \sin\eta, 0)$, and $\mathsf{O}_2$ is the rotation matrix
\begin{equation} \fl
{\sf O}_2 = \left(\!\!
\begin{array}{c@{\;}c@{\;}c}
 \sin\eta\cos\xi -\cos\eta\sin\zeta\sin\xi & -\sin\eta\sin\xi -\cos\eta\sin\zeta\cos\xi  & -\cos\eta\cos\zeta\\
-\cos\eta\cos\xi - \sin\eta\sin\zeta\sin\xi  & \cos\eta\sin\xi - \sin\eta\sin\zeta\cos\xi & -\sin\eta\cos\zeta\\
 \cos\zeta\sin\xi   &   \cos\zeta\cos\xi  &  -\sin\zeta
\end{array}\!\!
\right)\!.
\label{eq:lisarot}
\end{equation}
The functions $\eta = \Omega t + \eta_0$ and $\xi = -\Omega t + \xi_0$ that appear in this matrix represent the \emph{true anomaly} of the spacecraft formation and the relative phase of the spacecraft, respectively; $\zeta = - \pi/6$ sets the $60^\circ$ inclination of the triangle with respect to the ecliptic. The resulting orbits are shown in the left panel of Fig.\ \ref{fig:lisa}.

For simplicity, in this paper we set $\eta_0 = \xi_0 = 0$; however, to map these orbits to those used in the LISA Simulator \cite{PhysRevD.67.022001} and in the Mock LISA Data Challenges \cite{2006AIPC..873..619A,2006AIPC..873..625.,2007CQGra..24..551A,2008CQGra..25r4026B,2010CQGra..27h4009B}, we need to relabel spacecraft 1, 2, and 3 as 0, 2, and 1, and then set $\eta_0 = \kappa$, $\xi_0 = 3\pi/2 - \kappa + \lambda$, where $\kappa$ and $\lambda$ are the parameters defined below Eqs.\ (56) and (57) of Ref.\ \cite{PhysRevD.67.022001}.

The apparent motion in the spacecraft rest frame of a GW source that is located at ecliptic latitude $\beta$ and longitude $\lambda$, and that has principal polarization axes oriented along $\psi$ (see Sec.\ \ref{sec:gwsource}) can be derived by requiring that the geometric products that appear in Eq.\ \eqref{eq:estawahl} remain unchanged as the spacecraft move along their orbits. Specifically we require that
\begin{equation} \fl
\qquad \hat{n}^\mathrm{moving}_{ij} \cdot \hat{k}^\mathrm{true}_{ij} = 
(\mathsf{O}_2 \hat{n}^\mathrm{rest}_{ij}) \cdot \hat{k}^\mathrm{true}_{ij} =
\hat{n}^\mathrm{rest}_{ij} \cdot (\mathsf{O}^T_2 \hat{k}^\mathrm{true}_{ij}) =
\hat{n}^\mathrm{rest}_{ij} \cdot \hat{k}^\mathrm{moving}_{ij};
\end{equation}
by equating the components of $\hat{k}^\mathrm{moving}_{ij} \equiv -(\cos \lambda' \cos \beta', \sin \lambda' \cos \beta', \sin \beta')$ with those of $\mathsf{O}^T_2 \hat{k}^\mathrm{true}_{ij} \equiv -\mathsf{O}^T_2 \cdot (\cos \lambda \cos \beta, \sin \lambda \cos \beta, \sin \beta)$ we can then solve for $\beta'(t)$ and $\lambda'(t)$ as functions of $\beta$, $\lambda$, $\eta(t)$, and $\xi(t)$ (more precisely, we use the three equations to solve for $\cos \lambda'$, $\sin \lambda'$, and $\sin \beta'$). Some examples of the resulting apparent motion are shown in the right panel of Fig.\ \ref{fig:lisa}. To obtain the apparent variation $\psi'(t)$ of the polarization angle we need an additional equation that we can obtain from
\begin{equation} \fl
\qquad \hat{n}^\mathrm{moving}_{ij} \cdot 
[\mathsf{O}^\mathrm{true}_1 \mathsf{e}_+ (\mathsf{O}^\mathrm{true}_1)^T]
\cdot \hat{n}^\mathrm{moving}_{ij} =
\hat{n}^\mathrm{rest}_{ij} \cdot 
[\mathsf{O}^\mathrm{moving}_1 \mathsf{e}_+ (\mathsf{O}^\mathrm{moving}_1)^T]
\cdot \hat{n}^\mathrm{rest}_{ij},
\end{equation}
which implies
\begin{equation}
\mathrm{O}^\mathrm{moving}_1(\beta',\lambda',\psi') = \mathsf{O}^T_2(\eta,\xi) \cdot \mathrm{O}^\mathrm{rest}_1(\beta,\lambda,\psi).
\end{equation}
This is a matrix equation, but any two components that depend on $\cos \psi'$ and $\sin \psi'$ can be used to solve for $\psi'(\beta,\lambda,\psi)$ after replacing the functions found above for $\beta'$ and $\lambda'$.

\section{A fit to the averaged sensitivity for stationary and moving LISA-like detectors}
\label{app:snr2fit}

\begin{figure}
\flushright
\includegraphics[width=0.81\textwidth]{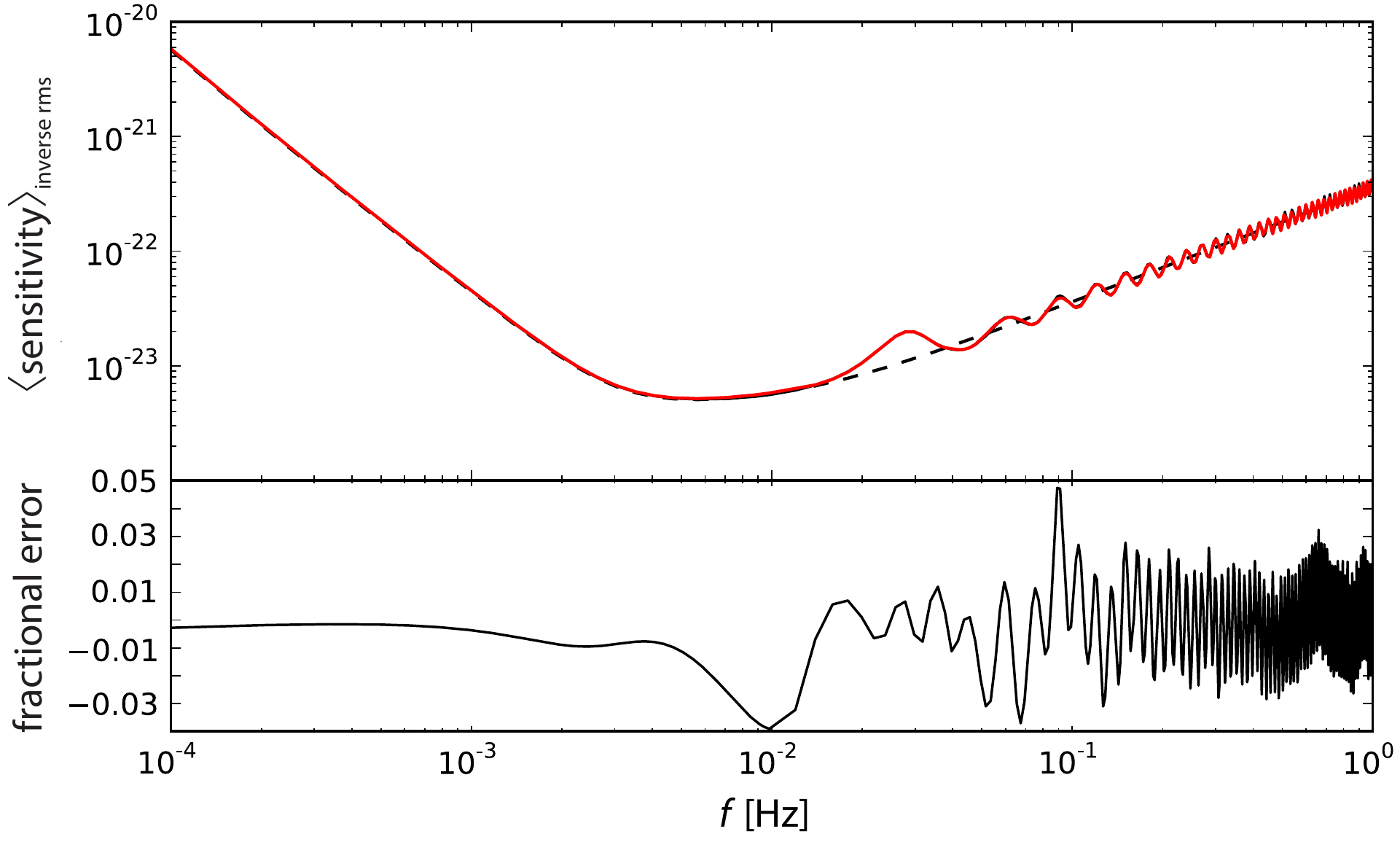}
\caption{\textbf{Top}: inverse-rms averaged sensitivity for an isotropic distribution of source geometries for the ``classic LISA'' configuration (black, solid), the standard fit of Eq.\ \eqref{simplefit1} (slightly improved by rescaling $S^\mathrm{op}$) (black, dashed), and the better fit of Eq.\ \eqref{improvedfit1} (red, solid). As usual, here we set $\mathrm{SNR}_\mathrm{thr} = 5$, $T = 1$ yr. 
\textbf{Bottom}: Fractional error of the improved fit.
\label{fig:snr2fit}}
\end{figure}

The LISA science requirements document \cite{lisasens} provides the standard approximation for the inverse-rms averaged effective-strain noise for LISA-like detectors,
\begin{equation} \fl
\qquad S^\mathrm{rms,std}_\mathrm{eff}(f) = \frac{5}{\sin^2 60^\circ} \bigg[ 1 + \bigg( \frac{2 f L}{0.41 \, c} \bigg)^{\!\!2} \bigg] \bigg( \frac{4 \, S^{\rm pm}(f)^{2} }{ (2\pi f )^4 } + S^{\rm op}(f)^2 \bigg) / L^2,
\label{simplefit1}
\end{equation}
where $60^\circ$ is the angle between the interferometer arms, $c$ is the speed of light, $L$ is the interferometer armlength ($5 \times 10^6$ km for LISA), and the optical-path and proof-mass noise components are given in Table \ref{tab:noises} (for this expression, one should use their values in physical units). Equation \eqref{simplefit1} does not include the Galactic confusion foreground. The sensitivity can then be obtained as $\mathrm{SNR}_\mathrm{thr} \sqrt{S^\mathrm{rms}_\mathrm{eff} / T}$.

Equation \eqref{simplefit1} does not model the oscillatory response-function behavior evident at frequencies larger than about $c / 2L$ ($30$ mHz for LISA). We can do better by replacing $S^{\rm op}$ with $0.964 \, S^{\rm op}$, and by multiplying Eq.\ \eqref{simplefit1} by the correction function
\begin{eqnarray} \fl
\qquad S_\mathrm{eff}^\mathrm{rms,fit} =
S_\mathrm{eff}^\mathrm{rms,std} \times 
\left\{ \begin{array}{ll}
	1 & f \le 0.015 {\rm ~Hz} \\ 
	a_0 + {\rm Skew} (f ; A, \xi, \alpha, \omega) & 0.015 < f \le 0.0375 {\rm ~Hz} \\
	a_1 \sin^2 \bigl( a_2 (f - a_3) \bigr) + a_4 & f \ge 0.0375 {\rm ~Hz}
	\end{array}\!\!, \right.
\label{improvedfit1}
\end{eqnarray}
where $a_0 = 0.9587$, $a_1 = 0.3319$, $a_2 = 104.7$, $a_3 = 0.01501$, and $a_4 = 0.8309$. The function ${\rm Skew}(f; A, \xi, \alpha, \omega)$ is the skew-normal distribution
\begin{equation} \fl
\qquad {\rm Skew} (f; A, \xi , \alpha , \omega) = \frac{ A}{ \omega \sqrt{2\pi} } \exp \left( - \frac{ (f - \xi)^2 }{ 2 \omega^2 } \right) \left[ 1 + {\rm erf} \left( \frac{ \alpha (f - \xi) }{ \sqrt{2} \omega } \right) \right],
\end{equation}
with $A = 0.009752$, $\alpha = -1.3908$, $\xi = 0.03093$, and $\omega = 0.006365$. Figure \ref{fig:snr2fit} shows the sensitivity curves resulting from this fit and from Eq.\ \eqref{simplefit1}, compared with the numerically derived inverse rms sensitivity (the yellow solid curve of Fig.\ \ref{fig:sensitivities1} and of the bottom panel of Fig.\ \ref{fig:sensitivities2}). The fit is accurate to better than $5\%$: the skew-normal distribution fits the first ``hump'' between $15$ and $37.5$ mHz to $1\%$, whereas the sinusoid fits the remaining oscillations to $3\%$.

\section*{References}

\bibliography{sensitivity}

\end{document}